\documentclass[english,twocolumn,superscriptaddress]{revtex4-1}
\usepackage[T1]{fontenc}
\usepackage[latin9]{inputenc}
\usepackage{color}
\usepackage{babel}
\usepackage{float}
\usepackage{amsmath}
\usepackage{amssymb}
\usepackage{graphicx}
\usepackage[unicode=true,pdfusetitle,
 bookmarks=true,bookmarksnumbered=false,bookmarksopen=false,
 breaklinks=false,pdfborder={0 0 1},backref=false,colorlinks=true]
 {hyperref}

\begin{document}
\title{Transverse momentum and multiplicity dependence of $\Lambda_{c}^{+}/D^{0}$
ratio in $pp$ collisions at $\sqrt{s}=13$ TeV}

\author{Jun Song }
\affiliation{School of Physical Science and Intelligent Engineering, Jining University,
Shandong 273155, China}

\author{Hai-hong Li}
\affiliation{School of Physical Science and Intelligent Engineering, Jining University,
Shandong 273155, China}

\author{Feng-lan Shao}
\email{shaofl@mail.sdu.edu.cn}
\affiliation{School of Physics and Physical Engineering, Qufu Normal University,
Shandong 273165, China}

\begin{abstract}
We apply an equal-velocity quark combination model to study the $\Lambda_{c}^{+}/D^{0}$
ratio in the range $p_{T}\lesssim10$ GeV/c in $pp$ collisions at
$\sqrt{s}=13$ TeV. We decompose the ratio into four parts which are
related to quark numbers, light-flavor quark $p_{T}$ spectrum, charm
quark $p_{T}$ spectrum, momentum correlation between light and charm
quarks, respectively. Their influence on $\Lambda_{c}^{+}/D^{0}$
ratio are individually studied. The curvature property of light-flavor
quark $p_{T}$ spectrum is found to be the main reason of the non-monotonic
$p_{T}$ dependence of $\Lambda_{c}^{+}/D^{0}$ ratio exhibited in
high multiplicity events. Moreover, the multiplicity dependence of
$\Lambda_{c}^{+}/D^{0}$ ratio as the function of $p_{T}$ is mainly
because of the multiplicity dependence of light-flavor quark $p_{T}$
spectrum. Using the light-flavor quark $p_{T}$ spectrum obtained
from experimental data of light-flavor hadrons and charm quark $p_{T}$
spectrum obtained from FONLL and/or PYTHIA calculations, the $p_{T}$
dependence of experimental data of $\Lambda_{c}^{+}/D^{0}$ ratio
in high multiplicity events and that in low multiplicity events in
$pp$ collisions at $\sqrt{s}=13$ TeV are reasonably understood.
\end{abstract}

\maketitle

\section{Introduction}

In recent years, LHC experiments provided rich experimental data of
hadron production in $pp$ collisions at LHC energies, from which
many new features of hadron production are found. For example, in
production of light-flavor hadrons, experiments found the enhancement
of baryon to meson ratio (such as $p/\pi$, $\Lambda/K_{s}^{0}$,
$\Omega/\phi$) at intermediate $p_{T}$ \citep{ALICE:2020jsh,Khachatryan:2011tm}
and the enhancement of multi-strange hadrons in high multiplicity
events \citep{ALICE:2017jyt}. We also find a quark number scaling
property for $p_{T}$ spectra of hadrons at mid-rapidity by further
analyzing experimental data of ALICE collaboration \citep{Song:2017gcz,Zhang:2018vyr,Li:2021nhq}.
In production of open heavy-flavor hadrons, LHC experiments found
the enhancement of $\Lambda_{c}^{+}/D^{0}$ in the low $p_{T}$ range
($p_{T}\lesssim10$ GeV/c) in $pp$ collisions at LHC energies in
comparison with measurements in $e^{+}e^{-}$ and $ep$ collisions
at early years \citep{Acharya:2017kfy,CMS:2019uws,ALICE:2020wla,ALICE:2020wla,ALICE:2021rzj}. 

The production enhancement of light-flavor baryons and, in particular,
heavy-flavor baryons such as $\Lambda_{c}^{+}$ attracts lots of theoretical
studies. Many new phenomenological mechanism and/or details at the
hadronization of final-parton system created in $pp$ collisions at
LHC energies, either in fragmentation framework \citep{Ortiz:2013yxa,Bierlich:2014xba,Bierlich:2015rha,Gieseke:2017clv,Christiansen:2015yqa,Fischer:2016zzs,Sjostrand:2017cdm}
or in quark (re-)combination framework, are proposed to explain these
new experimental data, which greatly enrich people's understandings
for the property of hadron production in high energy collisions. Recently,
ALICE collaboration report their precise measurement for the multiplicity
dependence of $\Lambda_{c}^{+}/D^{0}$ ratio in the low $p_{T}$ range
\citep{ALICE:2021npz} and the preliminary data $\Lambda_{c}^{+}/D^{0}$
ratio at small $p_{T}$ ($p_{T}<1$ GeV/c) \citep{Wilkinson:2022tyl,ALICE:2022ych}.
These newest data will further test the existing hadronization models
\citep{He:2019tik,Minissale:2020bif,Chen:2020drg,Song:2018tpv}. 

In this paper, we apply a quark combination model \citep{Song:2018tpv,Li:2017zuj,Song:2017gcz}
to study the multiplicity and $p_{T}$ dependence of $\Lambda_{c}^{+}/D^{0}$
ratio in $pp$ collisions at $\sqrt{s}=13$ TeV. The model applies
an equal-velocity combination (EVC) mechanism to approximately describe
the combination of constituent (anti-)quarks at hadronization. Taking
advantage of the analytic feature of EVC mechanism, we derive the
analytical formula of the ratio $\Lambda_{c}^{+}/D^{0}$ and study
several physical ingredients that influence the multiplicity and $p_{T}$
dependence of the ratio $\Lambda_{c}^{+}/D^{0}$. We demonstrate how
the property of light-flavor quark $p_{T}$ spectra and charm quark
$p_{T}$ spectra lead to, in an intuitive way, the non-monotonic $p_{T}$
dependence the ratio $\Lambda_{c}^{+}/D^{0}$. Finally, we compare
our theoretical calculations with the latest experimental data of
ALICE collaboration \citep{ALICE:2021npz}. 

The paper is organized as follows. Sec. \ref{sec:model} gives a brief
introduction to our quark combination model with EVC mechanism. In
Sec. \ref{sec:Lamc_D0_deri}, we derive the ratio $\Lambda_{c}^{+}/D^{0}$
and decompose it into four parts. In Secs. \ref{sec:Rnq}-\ref{sec:R_corr_cl},
we discuss the influence of these four parts, i.e., quark numbers,
light-flavor quark $p_{T}$ spectra, charm quark $p_{T}$ spectrum,
and $p_{T}$ correlation between charm and light-flavor (anti-)quarks,
on the ratio $\Lambda_{c}^{+}/D^{0}$. In Sec. \ref{sec:Ratio_LamC_D0_compare},
we compare our theoretical results of ratio $\Lambda_{c}^{+}/D^{0}$
with experimental data. In Sec. \ref{sec:summary}. we give the summary. 

\section{A brief introduction to EVC model\label{sec:model}}

In this paper, we apply a quark combination model proposed in previous
works \citep{Song:2018tpv,Li:2017zuj,Song:2017gcz} to study ratio
$\Lambda_{c}^{+}/D^{0}$. This model is inspired by the quark number
scaling property found from experimental data for $p_{T}$ spectra
of light-flavor hadrons at mid-rapidity in $pp$ and $p$-Pb collisions
at LHC energies \citep{Song:2017gcz,Gou:2017foe}. We have applied
the model to describe the yield and $p_{T}$ spectra of light-flavor
hadrons and those of single-charm hadrons in $pp$, $p$-Pb and AA
collision at RHIC and LHC energies, and found generally good agreement
with experimental data \citep{Song:2021ygg,Song:2021ojn,Song:2020kak,Li:2021nhq,Wang:2019fcg}.
In this section, we briefly introduce the model and, in particular,
the relevant physical approximations and parameters in the model which
may influence the ratio $\Lambda_{c}^{+}/D^{0}$. 

\subsection{general framework\label{subsec:EVC_model}}

We start from the inclusive momentum distribution of single-charm
hadron in general framework of quark combination mechanism 
\begin{align}
f_{M_{c\bar{l}}}(p) & =\int dp_{1}dp_{2}f_{c\bar{l}}(p_{1},p_{2})\,{\cal R}_{M_{c\bar{l}}}(p_{1},p_{2};p),\label{eq:fmc}\\
f_{B_{cll'}}(p) & =\int dp_{1}dp_{2}dp_{3}f_{cll'}(p_{1},p_{2},p_{3})\,{\cal R}_{B_{cll'}}(p_{1},p_{2},p_{3};p).\label{eq:fbc}
\end{align}
Here, $f_{c\bar{l}}(p_{1},p_{2})$ is the joint momentum distribution
of charm ($c$) quark and light anti-quark ($\bar{l}$). ${\cal R}_{M_{c\bar{l}}}(p_{1},p_{2};p)$
is the combination function denoting the probability density for the
given $c\bar{l}$ with momenta $p_{1}$, $p_{2}$ combining into a
meson $M_{c\bar{l}}$ with momentum $p$. It is similar for the baryon
formula.

In our model, we assume that the charm hadron is formed mainly by
the combination of charm quark with light-flavor (anti-)quarks with
equal velocity. In this approximation, the combination function is
simply the product of Dirac delta functions
\begin{align}
{\cal R}_{M_{c\bar{l}}}(p_{1},p_{2};p) & =\kappa_{M_{c\bar{l}}}\prod_{i=1}^{2}\delta(p_{i}-x_{i}p),\label{eq:rmc}\\
{\cal R}_{B_{cll'}}(p_{1},p_{2},p_{3};p) & =\kappa_{B_{cll'}}\prod_{i=1}^{3}\delta(p_{i}-x_{i}p).\label{eq:rbc}
\end{align}
 Here $\kappa_{M_{c\bar{l}}}$ and $\kappa_{B_{cll'}}$ are independent
of momentum but are dependent on quarks numbers due to the unitarity
constraint of hadronization. Momentum fraction reads as $x_{i}=m_{i}/(m_{1}+m_{2})$
in meson formula with momentum conservation constraint $x_{1}+x_{2}=1$
and $x_{i}=m_{i}/(m_{1}+m_{2}+m_{3})$ in baryon formula with momentum
conservation constraint $x_{1}+x_{2}+x_{3}=1$.

Substituting Eqs.~(\ref{eq:rmc}) and (\ref{eq:rbc}) into (\ref{eq:fmc})
and (\ref{eq:fbc}), we obtain 
\begin{align}
f_{M_{c\bar{l}}}(p) & =\kappa_{M_{c\bar{l}}}f_{c\bar{l}}(x_{1}p,x_{2}p),\label{eq:fpt_mcl_kappa}\\
f_{B_{cll'}}(p) & =\kappa_{B_{cll'}}f_{cll'}(x_{1}p,x_{2}p,x_{3}p).\label{eq:fpt_bcll_kappa}
\end{align}
For the integral of joint distribution of quarks, we have 
\begin{align}
\int dp_{1}dp_{2}f_{c\bar{l}}(p_{1},p_{2}) & =N_{c}N_{\bar{l}}\equiv N_{c\bar{l}}\\
\int dp_{1}dp_{2}dp_{3}f_{cll'}(p_{1},p_{2},p_{3}) & =N_{c}N_{ll'}\equiv N_{cll'}
\end{align}
where $N_{ll'}$ equals to $N_{l}N_{l'}$ as $l\neq l'$ and $N_{l}(N_{l}-1)$
as $l\neq l'$. Obviously, $N_{c\bar{l}}$ is the number of all $c\bar{l}$
pair and $N_{cll'}$ is the number of all possible $cll'$ combinations.
Under EVC constraint, the integrals become
\begin{align}
\int dpf_{c\bar{l}}(x_{1}p,x_{2}p) & \equiv\frac{N_{c\bar{l}}}{A_{c\bar{l}}},\label{eq:A_Mcl_def}\\
\int dpf_{cll'}(x_{1}p,x_{2}p,x_{3}p) & \equiv\frac{N_{cll'}}{A_{cll'}},\label{eq:A_Bcll_def}
\end{align}
where we use coefficient $A_{c\bar{l}}$ to denote the effect of equal-velocity
constraint to the effective number of $c\bar{l}$ pairs and $A_{cll'}$
to denote that to effective number of $cll'$ combinations.

Integrating Eqs. (\ref{eq:fpt_mcl_kappa}) and (\ref{eq:fpt_bcll_kappa})
over $p$, we obtain the number of charm hadrons
\begin{align}
N_{M_{c\bar{l}}} & =\kappa_{M_{c\bar{l}}}\frac{N_{c\bar{l}}}{A_{c\bar{l}}}=N_{c\bar{l}}\frac{\kappa_{M_{c\bar{l}}}}{A_{c\bar{l}}},\label{eq:N_Mcl}\\
N_{B_{cll'}} & =\kappa_{B_{cll'}}\frac{N_{cll'}}{A_{cll'}}=N_{cll'}\frac{\kappa_{B_{cll'}}}{A_{cll'}}.\label{eq:N_Bcll}
\end{align}
We see that $\frac{\kappa_{M_{c\bar{l}}}}{A_{c\bar{l}}}$ has the
meaning of the average probability of a $c\bar{l}$ pair forming a
meson, and it is similar for baryon term. 

If we neglect the contribution of multi-charm hadrons, the unitarity
of charm quark hadronization gives 
\begin{equation}
\sum_{M_{c\bar{l}}}N_{M_{c\bar{l}}}+\sum_{B_{cll'}}N_{B_{cll'}}=N_{c},
\end{equation}
which means
\begin{equation}
\sum_{M_{c\bar{l}}}N_{\bar{l}}\frac{\kappa_{M_{c\bar{l}}}}{A_{c\bar{l}}}+\sum_{B_{cll'}}N_{ll'}\frac{\kappa_{B_{cll'}}}{A_{cll'}}=1.
\end{equation}
 At charm quark hadronization, light-flavor quarks serve as the background
and their property, i.e., numbers $N_{l}$ and momentum distributions
$f_{l}(p)$ relating to $A$, can be freely changed. With this consideration,
we expect $\frac{\kappa_{M_{c\bar{l}}}}{A_{c\bar{l}}}\sim\frac{1}{N_{\bar{q}}}$
and $\frac{\kappa_{B_{cll'}}}{A_{cll'}}\sim\frac{1}{N_{q}^{2}}$ with
$N_{\bar{q}}=\sum_{\bar{l}}N_{\bar{l}}$ and $N_{q}=\sum_{l}N_{l}$
so that the unitarity constraint can be satisfied in an easy manner.
In this philosophy, we firstly introduce a dynamic parameter $R_{B/M}^{(c)}$
to denote the competition between the formation of charm baryon and
the formation of charm meson as a charm quark hadronizes. Then we
can write
\begin{align}
\sum_{M_{c\bar{l}}}N_{\bar{l}}\frac{\kappa_{M_{c\bar{l}}}}{A_{c\bar{l}}} & =\frac{1}{1+R_{B/M}^{(c)}},\\
\sum_{B_{cll'}}N_{ll'}\frac{\kappa_{B_{cll'}}}{A_{cll'}} & =\frac{R_{B/M}^{(c)}}{1+R_{B/M}^{(c)}}.
\end{align}
Following the spirit in our previous works \citep{Li:2017zuj,Song:2018tpv},
we can take the following parameterizations 
\begin{align}
\frac{\kappa_{M_{c\bar{l}}}}{A_{c\bar{l}}} & =C_{M_{c\bar{l}}}\frac{1}{N_{\bar{q}}}\frac{1}{1+R_{B/M}^{(c)}}=C_{M_{c\bar{l}}}\frac{1}{N_{c\bar{q}}}\frac{1}{1+R_{B/M}^{(c)}}N_{c},\label{eq:kappa_mcl}\\
\frac{\kappa_{B_{cll'}}}{A_{cll'}} & =C_{B_{cll'}}N_{iter,ll'}\frac{1}{N_{qq}}\frac{R_{B/M}^{(c)}}{1+R_{B/M}^{(c)}} \nonumber \\ 
    & =C_{B_{cll'}}N_{iter,ll'}\frac{1}{N_{cqq}}\frac{R_{B/M}^{(c)}}{1+R_{B/M}^{(c)}}N_{c},\label{eq:kappa_bcll}
\end{align}
which can be understood as follows. In the second equality, $N_{c}/(1+R_{B/M}^{(c)})$
denotes the number of charm quarks that will form charm mesons. $N_{c\bar{q}}$
equals to $N_{c}N_{\bar{q}}$ and denotes the number of all possible
$c\bar{q}$ pair. Then $N_{c}/(1+R_{B/M}^{(c)})/N_{c\bar{q}}$ denotes
the average probability of a $c\bar{q}$ pair forming a charm meson.
Because $c\bar{q}$ pair can form the charm meson with different total
angular moments, here we introduce a parameter $C_{M_{c\bar{l}}}$
to denote the probability of a $c\bar{l}$ pair forming a given state
$M_{c\bar{l}}$. Obviously, unitarity requires $\sum_{M}C_{M_{c\bar{l}}}=1$
where summation runs over all meson states with the same $c\bar{l}$
composition. It is similar for Eq.~(\ref{eq:kappa_bcll}) of the
baryon. Here, $N_{iter,ll'}$ is the permutation factor of $ll'$
and is take 2 for $l\neq l'$ and 1 for $l=l'$, respectively. We
expect $C_{M_{c\bar{l}}}$ and $C_{B_{cll'}}$ are stable. Because
$\sum_{M_{c\bar{l}}}N_{\bar{l}}C_{M_{c\bar{l}}}=\sum_{\bar{l}}N_{\bar{l}}=N_{\bar{q}}$
and $\sum_{B_{cll'}}C_{B_{cll'}}N_{iter,ll'}N_{ll'}=\sum_{ll'}N_{iter,ll'}N_{ll'}=N_{qq}$,
the unitarity of charm hadronization is naturally satisfied via $\sum_{M_{c\bar{l}}}N_{M_{c\bar{l}}}=\sum_{M_{c\bar{l}}}N_{c\bar{l}}\kappa_{M_{c\bar{l}}}/A_{c\bar{l}}=N_{c}/(1+R_{B/M}^{(c)})$
and $\sum_{B_{cll'}}N_{B_{cll'}}=\sum_{B_{cll'}}N_{cll'}\kappa_{B_{cll'}}/A_{cll'}=N_{c}R_{B/M}^{(c)}(1+R_{B/M}^{(c)})$.

Finally, numbers of charm hadrons are 
\begin{align}
N_{M_{c\bar{l}}} & =C_{M_{c\bar{l}}}\frac{1}{1+R_{B/M}^{(c)}}\frac{N_{c\bar{l}}}{N_{\bar{q}}},\\
N_{B_{cll'}} & =C_{B_{cll'}}\frac{R_{B/M}^{(c)}}{1+R_{B/M}^{(c)}}N_{iter,ll'}\frac{N_{cll'}}{N_{qq}},
\end{align}
and momentum distribution of charm hadrons are
\begin{align}
f_{M_{c\bar{l}}}(p) & =N_{M_{c\bar{l}}}f_{c\bar{l}}^{(n)}(x_{1}p,x_{2}p),\\
f_{B_{cll'}}(p) & =N_{B_{cll'}}f_{cll'}^{(n)}(x_{1}p,x_{2}p,x_{3}p),
\end{align}
where $f_{c\bar{l}}^{(n)}(x_{1}p,x_{2}p)$ and $f_{cll'}^{(n)}f_{cll'}^{(n)}(x_{1}p,x_{2}p,x_{3}p)$
are the normalized distributions under integral over $p$. 

\subsection{application to high-energy collisions\label{subsec:event_average}}

In this paper, we study the production of charm hadron at mid-rapidity
in $pp$ collisions at $\sqrt{s}=13$ TeV and compare our theoretical
results with experimental data at mid-rapidity. The quark momentum
distribution $dN/dp$ at rapidity $y=0$ in our model is reduced to
$f(p_{T})\equiv dN/dp_{T}$. The joint quark momentum distributions
are also reduced to $f_{c\bar{l}}(p_{T,c},p_{T,\bar{l}})$ and $f_{cll'}(p_{T,c},p_{T,l},p_{T,l'})$.
The rapidity densities of charm and light-flavor quarks are denoted
as $N_{l}$ and $N_{c}$ for convenience.

In experimental measurement, momentum spectra and yield densities
of hadrons are mainly reported via their averaged values in the selected
event class. We can extend the above formulas to follows experimental
statistics by re-defining quark momentum distributions $f_{c\bar{l}}(p_{T,c},p_{T,\bar{l}})$
and $f_{cll'}(p_{T,c},p_{T,l},p_{T,l'})$ and quark numbers $N_{l}$,
$N_{c}$ as these in the select event class. 

Finally, $p_{T}$ spectra of hadrons in our model are 

\begin{align}
f_{M_{c\bar{l}}}(p_{T}) & =\left\langle N_{M_{c\bar{l}}}\right\rangle f_{c\bar{l}}^{(n)}(x_{1}p_{T},x_{2}p_{T}),\\
f_{B_{cll'}}(p_{T}) & =\left\langle N_{B_{cll'}}\right\rangle f_{cll'}^{(n)}(x_{1}p_{T},x_{2}p_{T},x_{3}p_{T}),
\end{align}
with 
\begin{align}
\left\langle N_{M_{c\bar{l}}}\right\rangle  & =C_{M_{c\bar{l}}}\frac{1}{1+R_{B/M}^{(c)}}\left\langle \frac{N_{c\bar{l}}}{N_{\bar{q}}}\right\rangle ,\\
\left\langle N_{B_{cll'}}\right\rangle  & =C_{B_{cll'}}\frac{R_{B/M}^{(c)}}{1+R_{B/M}^{(c)}}N_{iter,ll'}\left\langle \frac{N_{cll'}}{N_{qq}}\right\rangle .
\end{align}

In the following studies, we also assume several symmetry property
for numbers and momentum distributions of quarks produced at mid-rapidity
in $pp$ collisions at LHC energy. We consider the iso-spin symmetry
between up and down quarks, i.e., $\left\langle N_{u}\right\rangle =\left\langle N_{d}\right\rangle $
and $f_{u}(p_{T})=f_{d}(p_{T})$, and also the charge conjugation
symmetry, i.e.~, $\left\langle N_{l}\right\rangle =\left\langle N_{\bar{l}}\right\rangle $
and $f_{l}(p_{T})=f_{\bar{l}}(p_{T})$. These symmetry assumptions
will greatly simplify our theoretical expressions.

\section{$p_{T}$ dependence of ratio $\Lambda_{c}^{+}/D^{0}$\label{sec:Lamc_D0_deri}}

According to Eqs.~(\ref{eq:fpt_mcl_kappa}) and (\ref{eq:fpt_bcll_kappa}),
$p_{T}$ spectra of $\Lambda_{c}^{+}$ and $D_{0}$ are 
\begin{align}
f_{\Lambda_{c}}(p_{T}) & =\kappa_{\Lambda_{c}}f_{cud}\left(p_{T,c},p_{T,u},p_{T,d}\right),\\
f_{D^{0}}(p_{T}) & =\kappa_{D^{0}}f_{c\bar{u}}\left(x_{c}^{'}p_{T},x_{\bar{u}}^{'}p_{T}\right),
\end{align}
where $x_{u}=x_{d}=m_{u}/(2m_{u}+m_{c})$, $x_{c}=m_{c}/(2m_{u}+m_{c})$
for the baryon and $x_{\bar{u}}^{'}=m_{u}/(m_{u}+m_{c})$, $x_{c}^{'}=m_{c}/(m_{c}+m_{u})$
for the meson. 

We express the joint distribution functions of (anti-)quarks as 
\begin{align}
    &f_{cud}\left(p_{T,c},p_{T,u},p_{T,d}\right) \label{eq:fudc_decomp}\\
    &=f_{c}(p_{T,c})f_{u}(p_{T,u})f_{d}(p_{T,d})\left[1+\mathcal{C}_{cud}\left(p_{T,c},p_{T,u},p_{T,d}\right)\right], \nonumber \\
    &f_{c\bar{u}}\left(p_{T,c},p_{T,\bar{u}}\right) =f_{c}(p_{T,c})f_{\bar{u}}(p_{T,\bar{u}})\left[1+\mathcal{C}_{c\bar{u}}\left(p_{T,c},p_{T,\bar{u}}\right)\right],\label{eq:fuc_decomp}
\end{align}
where $f_{q_{i}}(p_{T})$ is inclusive distribution of $q_{i}$-flavor
quarks and $\mathcal{C}_{cud}\left(p_{T,c},p_{T,u},p_{T,d}\right)$
is the correlation term. By rewriting $f_{q_{i}}(p_{T})=N_{q_{i}}f_{q_{i}}^{(n)}(p_{T})$
with $f_{q_{i}}^{(n)}(p_{T})$ being the normalized distribution,
we can further write the joint distribution as
\begin{widetext}
\begin{align}
f_{cud}\left(p_{T,c},p_{T,u},p_{T,d}\right) & =N_{cud}f_{c}^{(n)}(p_{T,c})f_{u}^{(n)}(p_{T,u})f_{d}^{(n)}(p_{T,d})\left[1+\mathcal{C}_{cud}\left(p_{T,c},p_{T,u},p_{T,d}\right)\right],\label{eq:fudc_fin_decomp}\\
f_{c\bar{u}}\left(p_{T,c},p_{T,\bar{u}}\right) & =N_{c}N_{\bar{u}}f_{c}^{(n)}(p_{T,c})f_{\bar{u}}^{(n)}(p_{T,\bar{u}})\left[1+\mathcal{C}_{c\bar{u}}\left(p_{T,c},p_{T,\bar{u}}\right)\right].\label{eq:fuc_fin_decomp}
\end{align}

With the above form of $f_{cud}\left(p_{T,c},p_{T,u},p_{T,d}\right)$
and $f_{\bar{u}c}\left(p_{T,c},p_{T,\bar{u}}\right)$, we can calculate
$p_{T}$ spectra of $\Lambda_{c}^{+}$ and $D_{0}$ by formulas in
Sec. \ref{subsec:EVC_model} and then consider the event average convention
in Sec. \ref{subsec:event_average}, and finally we get the ratio
\begin{align}
\frac{f_{\Lambda_{c}}(p_{T})}{f_{D_{0}}(p_{T})} & =R_{B/M}^{(c)}\frac{C_{\Lambda_{c}}}{C_{D^{0}}}N_{iter,ud}\frac{\left\langle \frac{N_{cud}}{N_{qq}}\right\rangle }{\left\langle \frac{N_{c\bar{u}}}{N_{\bar{q}}}\right\rangle }\frac{A_{\Lambda_{c}}}{A_{D_{0}}}\frac{f_{c}^{(n)}(x_{c}p_{T})}{f_{c}^{(n)}(x_{c}^{'}p_{T})}\frac{\left[f_{u}^{(n)}(x_{u}p_{T})\right]^{2}}{f_{u}^{(n)}(x_{u}^{'}p_{T})}\frac{1+\mathcal{C}_{cud}\left(x_{c}p_{T},x_{u}p_{T},x_{d}p_{T}\right)}{1+\mathcal{C}_{\bar{u}c}\left(x_{c}^{'}p_{T},x_{u}^{'}p_{T}\right)}\nonumber \\
 & =R^{(N_{q_{i}})}\left[\frac{A_{\Lambda_{c}}}{A_{D_{0}}}R_{\Delta x_{c}}^{(c)}\left(p_{T}\right)R_{\Delta x_{l}}^{(l)}\left(p_{T}\right)R_{corr}^{(cl)}\left(p_{T}\right)\right].\label{eq:eq:ratio_LambdaC_D0_decomp}
\end{align}
\end{widetext}
 Here, we have used the symmetry property $f_{u}^{(n)}(p_{T})=f_{d}^{(n)}(p_{T})=f_{\bar{u}}^{(n)}(p_{T})=f_{\bar{d}}^{(n)}(p_{T})$
in the mid-rapidity range at LHC energies. In the second line, we
split the ratio into several parts 
\begin{align}
R^{(N_{q_{i}})} & \equiv2R_{B/M}^{(c)}\frac{C_{\Lambda_{c}}\left\langle N_{cud}/N_{qq}\right\rangle }{C_{D_{0}}\left\langle N_{c\bar{u}}/N_{\bar{q}}\right\rangle },\label{eq:R_Nqi}\\
R_{\Delta x_{c}}^{(c)}\left(p_{T}\right) & \equiv\frac{f_{c}^{(n)}(x_{c}p_{T})}{f_{c}^{(n)}(x_{c}^{'}p_{T})},\label{eq:R_dxc}\\
R_{\Delta x_{l}}^{(l)}\left(p_{T}\right) & \equiv\frac{\left[f_{u}^{(n)}(x_{u}p_{T})\right]^{2}}{f_{u}^{(n)}(x_{u}^{'}p_{T})},\label{eq:R_dxu}\\
R_{corr}^{(cl)}\left(p_{T}\right) & =\frac{1+\mathcal{C}_{cud}\left(x_{c}p_{T},x_{u}p_{T},x_{d}p_{T},\right)}{1+\mathcal{C}_{c\bar{u}}\left(x_{c}^{'}p_{T},x_{u}^{'}p_{T}\right)}\label{eq:R_cl_corr}
\end{align}
which reflect the influence of different physical ingredients on the
ratio $\Lambda_{c}^{+}/D^{0}$.

\section{The property of $R^{(N_{q_{i}})}$\label{sec:Rnq}}

With Eqs.~(\ref{eq:A_Mcl_def}), (\ref{eq:A_Bcll_def}) and (\ref{eq:fudc_fin_decomp}),
we can see that 
\begin{align}
    A_{\Lambda_{c}}\int dp_{T}&\left[f_{u}^{(n)}(x_{u}p_{T})\right]^{2}f_{c}^{(n)}(x_{c}p_{T})\nonumber \\
    &\times \left[1+\mathcal{C}_{cud}\left(x_{c}p_{T},x_{u}p_{T},x_{d}\text{\ensuremath{p_{T}}}\right)\right]  =1,\\
    A_{D_{0}}\int dp_{T}& f_{c}^{(n)}(x_{c}^{'}p_{T})f_{\bar{u}}^{(n)}(x_{u}^{'}p_{T})\nonumber \\
    &\times \left[1+\mathcal{C}_{c\bar{u}}\left(x_{c}^{'}p_{T},x_{u}^{'}p_{T}\right)\right]  =1,
\end{align}
which means that terms in square bracket in Eq.~(\ref{eq:eq:ratio_LambdaC_D0_decomp}),
as a global quantity, mainly influences the shape of the $\Lambda_{c}^{+}/D^{0}$
ratio but not its global magnitude. Therefore, $R^{(N_{q_{i}})}$
in Eq.~(\ref{eq:eq:ratio_LambdaC_D0_decomp}) plays the role of controlling
the global magnitude of $\Lambda_{c}^{+}/D^{0}$ ratio. In this section,
we discuss the property of $R^{(N_{q_{i}})}$ defined in Eq.~(\ref{eq:R_Nqi}). 

We take 
\begin{align}
\left\langle \frac{N_{udc}}{N_{qq}}\right\rangle  & \approx\frac{\left\langle N_{u}\right\rangle ^{2}\left\langle N_{c}\right\rangle }{\left\langle N_{q}\right\rangle ^{2}},\\
\left\langle \frac{N_{c\bar{u}}}{N_{\bar{q}}}\right\rangle  & \approx\frac{\left\langle N_{\bar{u}}\right\rangle \left\langle N_{c}\right\rangle }{\left\langle N_{\bar{q}}\right\rangle }.
\end{align}
Here, we use the approximations $\left\langle N_{\bar{u}}\right\rangle =\left\langle N_{\bar{d}}\right\rangle =\left\langle N_{u}\right\rangle =\left\langle N_{d}\right\rangle $
and $\left\langle N_{\bar{q}}\right\rangle =\left\langle N_{q}\right\rangle $
in the mid-rapidity range at LHC energies. We neglect quark number
correlations between different flavors by considering the small off-diagonal
susceptibilities of quark flavors in Lattice-QCD calculations \citep{Ding:2015fca}. 

In order to obtain intuitive expression, we define a strangeness suppression
factor 
\begin{equation}
\lambda_{s}=\frac{\left\langle N_{\bar{s}}\right\rangle }{\left\langle N_{\bar{u}}\right\rangle }.
\end{equation}
By noticing that $\left\langle N_{q}\right\rangle =\left\langle N_{u}\right\rangle +\left\langle N_{d}\right\rangle +\left\langle N_{s}\right\rangle =(2+\lambda_{s})\left\langle N_{u}\right\rangle $,
we obtain 
\begin{equation}
R^{\left(N_{q_{i}}\right)}=2R_{B/M}^{(c)}\frac{C_{\Lambda_{c}}}{C_{D^{0}}}\frac{1}{2+\lambda_{s}}.
\end{equation}
The branch fraction parameter $C_{\Lambda_{c}}$ and $C_{D^{0}}$
directly influence $R^{\left(N_{q_{i}}\right)}$. However, experimental
data of $\Lambda_{c}^{+}$ and $D^{0}$ usually contain the contribution
of strong and electromagnetic decays of other single-charm hadrons.
Considering the decay contributions of $\Sigma_{c}^{++,+,0}$ and
$\Sigma_{c}^{*++,+,0}$, the yield of final-state $\Lambda_{c}^{+}$
is 
\begin{align}
N_{\Lambda_{c}^{+}}^{(final)}= & \left(C_{\Lambda_{c}}+C_{\Sigma_{c}^{+}}\right)N_{iter,ud}\left\langle \frac{N_{udc}}{N_{qq}}\right\rangle \nonumber \\
    & +\left(C_{\Sigma^{++}}+C_{\Sigma^{*++}}\right)N_{iter,uu}\left\langle \frac{N_{uuc}}{N_{qq}}\right\rangle \nonumber \\
 & +\left(C_{\Sigma^{0}}+C_{\Sigma^{*0}}\right)N_{iter,dd}\left\langle \frac{N_{ddc}}{N_{qq}}\right\rangle \nonumber \\
 & =\left\langle \frac{N_{uuc}+2N_{udc}+N_{ddc}}{N_{qq}}\right\rangle \nonumber \\
    & \approx\frac{4}{\left(2+\lambda_{s}\right)^{2}}\left\langle N_{c}\right\rangle .\label{eq:C_Lambda_final}
\end{align}
Because the measured $D^{0}$ contains the decay contribution of $D^{*0,+}$,
the yield of final-state $D^{0}$ is 
\begin{align}
    & N_{D^{0}}^{(final)} \nonumber \\
    &=  C_{D^{0}}\left\langle \frac{N_{c\bar{u}}}{N_{\bar{q}}}\right\rangle +C_{D^{*0}}\left\langle \frac{N_{c\bar{u}}}{N_{\bar{q}}}\right\rangle +\mathcal{B}_{D^{*+}\to D^{0}}C_{D^{*+}}\left\langle \frac{N_{c\bar{d}}}{N_{\bar{q}}}\right\rangle \nonumber \\
 & =\left(1+\mathcal{B}_{D^{*+}\to D^{0}}C_{D^{*+}}\right)\left\langle \frac{N_{c\bar{u}}}{N_{\bar{q}}}\right\rangle \nonumber \\
 & \approx\left(1+\mathcal{B}_{D^{*+}\to D^{0}}C_{D^{*+}}\right)\frac{1}{2+\lambda_{s}}\left\langle N_{c}\right\rangle \label{eq:C_D0_final}
\end{align}
with decay branch ratio $\mathcal{B}_{D^{*+}\to D^{0}}=0.677$ from
particle data group \citep{Agashe:2014kda}. The branch fraction parameter
$C_{D^{*+}}$ is about 0.6 since the ratio $D^{*0}/D^{0}$ is about
0.43 \citep{Acharya:2019mgn}. With Eqs.~(\ref{eq:C_Lambda_final})
and (\ref{eq:C_D0_final}) we obtain $R^{\left(N_{q_{i}}\right)}$
for final-state $\Lambda_{c}^{+}/D^{0}$ ratio 
\begin{align}
R^{\left(N_{q_{i}},final\right)} & =R_{B/M}^{(c)}\frac{1}{1+0.677C_{D^{*+}}}\frac{4}{2+\lambda_{s}}\nonumber \\
 & =R_{B/M}^{(c)}\frac{2.85}{2+\lambda_{s}}.\label{eq:R_Nqi_final}
\end{align}
According to our estimations in previous work \citep{Zhang:2018vyr,Gou:2017foe},
strangeness suppression factor $\lambda_{s}$ in low multiplicity
events is about 0.3 and that in high multiplicity events is about
0.36 in $pp$ collisions at LHC energies. This change magnitude of
$\lambda_{s}$ causes little influence on $R^{\left(N_{q_{i}}\right)}$
through the term $2+\lambda_{s}$ and therefore contribute little
multiplicity dependence of $\Lambda_{c}^{+}/D^{0}$ ratio. 

$R_{B/M}^{(c)}$ is a dynamic parameter in our model that denotes
the production competition between baryon formation via combination
with two light-flavor quarks and meson formation via combination with
an anti-quark at a charm quark hadronization. In the environment of
abundant light-flavor quarks and antiquarks, charm quark has sufficient
chance to interact with surrounding light-flavor quarks and antiquarks,
and the baryon to meson production competition is sufficient and therefore
we expect $R_{B/M}^{(c)}$ should be saturated. 

However, in the low multiplicity events this baryon-to-meson production
competition is not sufficient due to the restricted numbers of light-flavor
quarks and/or anti-quarks. For example, in event multiplicity class
X with mid-rapidity $\left\langle dN_{ch}/d\eta\right\rangle =2.52$
in $pp$ collisions at $\sqrt{s}=$ 13 TeV \citep{ALICE:2019avo},
the up quark number density is $\left\langle dN_{u}/dy\right\rangle =1.48$
at mid-rapidity according to our previous study of multiplicity dependence
of light-flavor hadron production \citep{Zhang:2018vyr}. Events with
a charm quark and up/down quark with numbers $(N_{u},N_{d})=(2,0),(0,2)$
can not form $\Lambda_{c}^{+}$. Events with $(N_{u},N_{d})=(2,1)$
only provide one effective $ud$ pair for $\Lambda_{c}^{+}$ formation
but can provide three light-flavor antiquarks for charm meson formation.
Therefore, the formation of $\Lambda_{c}^{+}$ in these events is
suppressed relative to $D$ mesons. The fraction of these events should
not be too small due to the small values of $\left\langle dN_{u,d}/dy\right\rangle $
in low multiplicity events. Therefore, we can expect the suppression
of $R_{B/M}^{(c)}$ and corresponding $\Lambda_{c}^{+}/D^{0}$ ratio
to a certain extent in low multiplicity events in $pp$ collisions
at $\sqrt{s}=$ 13 TeV. 

The exact magnitude of such suppression is dependent on the property
of quark number distribution event-by-event, which is little known
at present. Here, we take the Poisson distribution as an example to
roughly estimate the magnitude of this suppression. We assume the
joint distribution of quark numbers as 
\begin{equation}
P(N_{u},N_{d})=\textrm{Poi}(N_{u})\textrm{Poi}(N_{d})
\end{equation}
with mean number $\left\langle N_{u}\right\rangle =\left\langle N_{d}\right\rangle $.
The number of $\bar{u}$ and that of $\bar{d}$ are taken to be $N_{\bar{u}}=N_{u}$
and $N_{\bar{d}}=N_{d}$, respectively. That is, we consider the pair
production in each event, for simplicity. For events with V0M multiplicity
classes IX and X with $\left\langle dN_{ch}/d\eta\right\rangle =2.52,4.64$
\citep{ALICE:2019avo}, the $u$ quark number is about $\left\langle N_{u}\right\rangle =1.5-2.5$
according to our previous study of light-flavor hadrons in $pp$ at
$\sqrt{s}=13$ TeV \citep{Zhang:2018vyr}. The fraction of events
that can form $D^{0}$ but not $\Lambda_{c}^{+}$ is 
\begin{equation}
P^{'}=\frac{1}{P_{nor}}\sum_{N_{u}=1}^{\infty}P(N_{u},0)=0.08-0.18
\end{equation}
with $P_{nor}=\sum_{N_{u},N_{d}}P(N_{u},N_{d})-P(0,0)$. In addition,
the fraction of events that can form $D^{0}$ but not $\Sigma_{c}^{0}$
or $\Sigma_{c}^{*0}$ which finally decay into $\Lambda_{c}^{+}$
is 
\begin{equation}
P^{''}=\frac{1}{P_{nor}}\sum_{N_{u}=1}^{\infty}P(N_{u},1)=0.19-0.27.
\end{equation}
The fraction of events that can not form $\Sigma_{c}^{++}$ or $\Sigma_{c}^{*++}$
is also $P^{''}$. These estimation therefore indicates a magnitude
of about 20\% suppression for $R_{B/M}^{(c)}$ in low multiplicity
events in comparison with that in high multiplicity events where the
above suppression is absent. In our previous works in study of single-charm
hadrons in $pp$ and $p$Pb collisions at LHC energies \citep{Li:2017zuj,Song:2018tpv},
we use $R_{B/M}^{(c)}=0.425\pm0.025$. If we take this value as the
possibly saturated value, we estimate that $R_{B/M}^{(c)}$ in low
multiplicity events is about 0.34. Therefore, we expect 
\begin{align}
    R^{\left(N_{q_{i}},final\right)} & =R_{B/M}^{(c)}\frac{2.85}{2+\lambda_{s}}\nonumber \\
    &\approx\begin{cases}
0.51 & \textrm{in high \ensuremath{\left\langle dN_{ch}/d\eta\right\rangle }events}\\
0.42 & \textrm{in low \ensuremath{\left\langle dN_{ch}/d\eta\right\rangle }events}
\end{cases}.\label{eq:R_Nqi_estimation}
\end{align}

\section{The property of $R_{\Delta x_{l}}^{(l)}\left(p_{T}\right)$\label{sec:R_dt_l}}

Because $m_{c}\approx5m_{u}$, the momentum fraction of up/down quark
$x_{u}^{'}=m_{u}/(m_{c}+m_{u})=0.167$ in $D^{0}$ is close to $x_{u}=m_{u}/(m_{c}+2m_{u})\approx0.143$
in $\Lambda_{c}^{+}$. The difference between them is quite small
\begin{equation}
\Delta x_{u} =x_{u}^{'}-x_{u}=\frac{m_{u}^{2}}{(m_{c}+m_{u})(m_{c}+2m_{u})}\approx0.024.
\end{equation}
In the studied range $p_{T}\lesssim8$ GeV/c, the momentum difference
$\Delta x_{u}p_{T}\lesssim0.2$ GeV/c is also small. In order to simplify
$R_{\Delta x_{l}}^{(l)}\left(p_{T}\right)$, we can take Taylor expansion
for $f_{u}(x_{u}p_{T})$ 
\begin{align}
f_{u}^{(n)}\left(x_{u}p_{T}\right) 
 & =f_{u}^{(n)}\left(x_{u}^{'}p_{T}\right)\Bigg[1-\frac{f_{u}^{(n)'}\left(x_{u}^{'}p_{T}\right)}{f_{u}^{(n)}\left(x_{u}^{'}p_{T}\right)}\Delta x_{u}p_{T} \nonumber \\
    &+\frac{1}{2}\frac{f_{u}^{(n)''}\left(x_{u}^{'}p_{T}\right)}{f_{u}^{(n)}\left(x_{u}^{'}p_{T}\right)}\left(\Delta x_{u}p_{T}\right)^{2}+....\Bigg],
\end{align}
where the contribution of perturbative terms are quite small, i.e.,~about
$0.01\sim0.07$ as $p_{T}<8$ GeV/c. Therefore, the light-flavor part
$R_{\Delta x_{l}}^{(l)}\left(p_{T}\right)$ becomes
\begin{align}
    &R_{\Delta x_{l}}^{(l)}\left(p_{T}\right)  =\frac{\left[f_{u}^{(n)}(x_{u}p_{T})\right]^{2}}{f_{u}^{(n)}(x_{u}^{'}p_{T})} &\nonumber \\
 &=f_{u}^{(n)}\left(x_{u}^{'}p_{T}\right)\Bigg[1-\frac{f_{u}^{(n)'}\left(x_{u}^{'}p_{T}\right)}{f_{u}^{(n)}\left(x_{u}^{'}p_{T}\right)}\Delta x_{u}p_{T} \nonumber \\
    &\hspace{70pt}+\frac{f_{u}^{(n)''}\left(x_{u}^{'}p_{T}\right)}{f_{u}^{(n)}\left(x_{u}^{'}p_{T}\right)}\left(\Delta x_{u}p_{T}\right)^{2}+....\Bigg]^{2}\nonumber \\
 & =f_{u}^{(n)}\left(x_{u}^{'}p_{T}\right)\left[1-2\frac{f_{u}^{(n)'}\left(x_{u}^{'}p_{T}\right)}{f_{u}^{(n)}\left(x_{u}^{'}p_{T}\right)}\Delta x_{u}p_{T}+....\right]\nonumber \\
 & \approx f_{u}^{(n)}\left((x_{u}^{'}-2\Delta x_{u})p_{T}\right).\label{eq:R_dxl_approx}
\end{align}
We see that $R_{\Delta x_{l}}^{(l)}\left(p_{T}\right)$ directly
depends on the property of $p_{T}$ spectrum of up/down quarks. 

\begin{figure*}[!htpb]
\centering{}
    \includegraphics[width=0.75\linewidth]{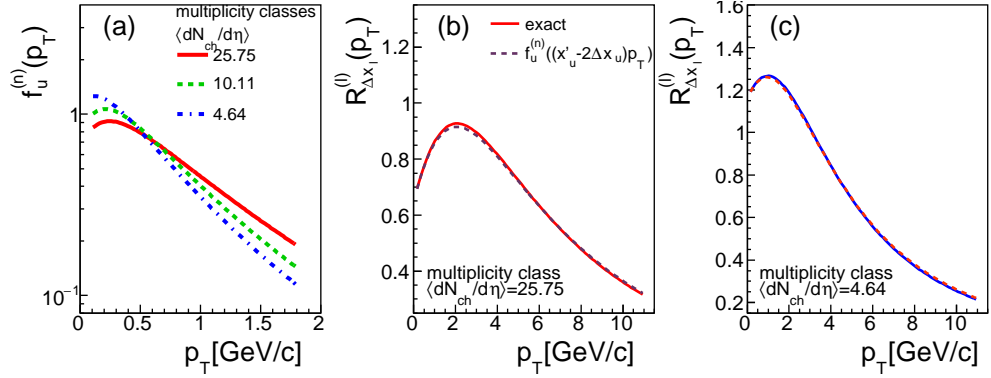}
    \caption{The normalized $p_{T}$ spectra of up quarks in three multiplicity classes in $pp$ collisions at $\sqrt{s}=13$ TeV (a), the calculated $R_{\Delta x_{l}}^{(l)}\left(p_{T}\right)$ in multiplicity class I with $\left\langle dN_{ch}/d\eta\right\rangle =25.75$ at mid-rapidity (b) and that in multiplicity class IX with $\left\langle dN_{ch}/d\eta\right\rangle =$4.64 (c). The solid lines in (b) and (c) are directly calculated from the definition by Eq.~(\ref{eq:R_dxu}) and the dashed lines are the approximations by Eq.~(\ref{eq:R_dxl_approx}). \label{fig:fu_Rl_dxl}}
\end{figure*}

In Fig.~\ref{fig:fu_Rl_dxl}(a), we show the normalized $p_{T}$
spectra of up quarks $f_{u}^{(n)}(p_{T})$ in the low $p_{T}$ range
(i.e., $p_{T,u}\lesssim2$ GeV/c) in three multiplicity classes in
$pp$ collisions at $\sqrt{s}=13$ TeV, which are obtained in previous
work \citep{Zhang:2018vyr} by fitting experimental data for $p_{T}$
spectra of proton at mid-rapidity using our EVC model. We see that
$f_{u}^{(n)}(p_{T})$ is different in three multiplicity classes.
In Fig.~\ref{fig:fu_Rl_dxl}(b), we show the property of $R_{\Delta x_{l}}^{(l)}\left(p_{T}\right)$
in high multiplicity class I with $\left\langle dN_{ch}/d\eta\right\rangle =25.75$
at mid-rapidity. Here, the V0M multiplicity classes are defined in
Ref.\citep{ALICE:2019avo}. The solid line denotes the result of direct
calculation of $R_{\Delta x_{l}}^{(l)}\left(p_{T}\right)$ by the
definition Eq.~(\ref{eq:R_dxu}) and the dashed line is the approximation
Eq.~(\ref{eq:R_dxl_approx}). We see that the approximation is very
good. We further see that $R_{\Delta x_{l}}^{(l)}\left(p_{T}\right)$
increases with $p_{T}$ in the range $0<p_{T}\lesssim3$ GeV/c and
decreases with $p_{T}$ at larger $p_{T}$. This property will cause
the non-monotonic $p_{T}$ dependence of ratio $\Lambda_{c}^{+}/D^{0}$
which is exhibited in experimental data in high multiplicity class.
We emphasize that the property of $f_{u}^{(n)}(p_{T})$ actually determine
the $p_{T}$ dependence of the ratio $\Lambda_{c}^{+}/D^{0}$ to a
large extent. In Fig.~\ref{fig:fu_Rl_dxl}(c), we show the property
of $R_{\Delta x_{l}}^{(l)}\left(p_{T}\right)$ in low multiplicity
class IX with $\left\langle dN_{ch}/d\eta\right\rangle =4.64$ at
mid-rapidity. We see that $R_{\Delta x_{l}}^{(l)}\left(p_{T}\right)$
only has a small shoulder structure in the range $p_{T}\lesssim1$
GeV/c and then decrease with $p_{T}$. This is the direct consequence
of $f_{u}^{(n)}(p_{T})$ in the low multiplicity class which exhibits
less \textbf{thermal} behavior in the low $p_{T}$ range, see the
dot-dashed line in Fig.~\ref{fig:fu_Rl_dxl}(a). Comparing Fig.~\ref{fig:fu_Rl_dxl}
(b) and (c), we emphasize that the multiplicity dependence of light-flavor
quark spectrum $f_{u}^{(n)}(p_{T})$ will lead to the significant
multiplicity dependence of $\Lambda_{c}^{+}/D^{0}$ ratio as a function
of $p_{T}$. 

\section{The property of $R_{\Delta x_{c}}^{(c)}\left(p_{T}\right)$\label{sec:R_dt_c}}

Because the difference between momentum fraction of charm quarks in
$D^{0}$ and that in $\Lambda_{c}^{+}$ 
\begin{equation}
\Delta x_{c} =x_{c}^{'}-x_{c}=\frac{m_{c}m_{u}}{(m_{c}+m_{u})(m_{c}+2m_{u})}\approx0.12
\end{equation}
is not quite small, the momentum difference is $\Delta x_{c}p_{T}\lesssim1$
GeV/c in the studied range $p_{T}\lesssim10$ GeV/c. Taylor expansion
is not generally effective to simplify $R_{\Delta x_{c}}^{(c)}\left(p_{T}\right)=f_{c}^{(n)}(x_{c}p_{T})/f_{c}^{(n)}(x'_{c}p_{T})$.
 Therefore, we take the specific form of $f_{c}^{(n)}(p_{T})$ and
directly evaluate $R_{\Delta x_{c}}^{(c)}\left(p_{T}\right)$. 

In practice, we take the following parameterization for $p_{T}$ spectrum
of charm quarks 
\begin{equation}
f_{c}^{(n)}\left(p_{T}\right)=\mathcal{N}p_{T}^{b}\left(1+\frac{\sqrt{p_{T}^{2}+M^{2}}-M}{nc}\right)^{-n},\label{eq:fcpt_para}
\end{equation}
where $b$, $n$, $M$, $c$ are parameters with positive values and
$\mathcal{N}$ is the normalization coefficient. This function can
usually well fit the $p_{T}$ spectrum of charm quarks obtained from
theoretical method as well as these of single-charm hadrons measured
by experiments. Substituting Eq.~(\ref{eq:fcpt_para}) into (\ref{eq:R_dxc}),
we obtain
\begin{align}
R_{\Delta x_{c}}^{(c)}\left(p_{T}\right) & =\frac{f_{c}^{(n)}(x_{c}p_{T})}{f_{c}^{(n)}(x_{c}^{'}p_{T})}\nonumber \\
 & =\left(\frac{x_{c}}{x_{c}^{'}}\right)^{b}\left(\frac{nc-M+\sqrt{\left(x_{c}^{'}p_{T}\right)^{2}+M^{2}}}{nc-M+\sqrt{\left(x_{c}p_{T}\right)^{2}+M^{2}}}\right)^{n},\label{eq:R_dxlc_appro}
\end{align}
which is slowly increasing with $p_{T}$ due to the positive exponent
$n$ and $x_{c}^{'}>x_{c}$. 

\begin{figure*}[!htpb]
\begin{centering}
\includegraphics[width=0.75\linewidth]{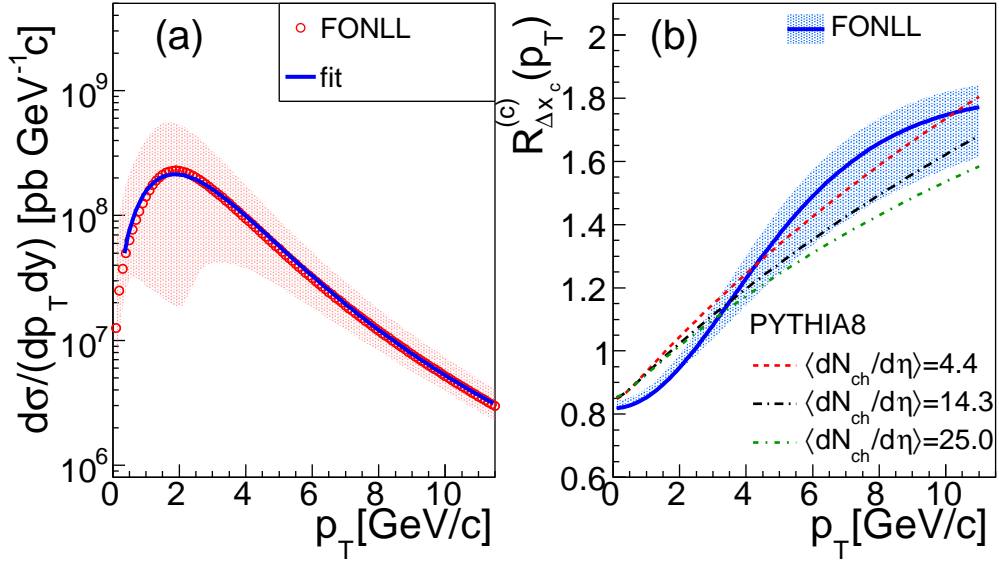}
    \caption{Panel (a): the differential cross-section of charm quarks at mid-rapidity in $pp$ collisions at $\sqrt{s}=13$ TeV calculated by FONLL method.  The shadow area denotes scale uncertainties. The solid line shows the fit of central points of theoretical calculations with Eq.~(\ref{eq:fcpt_para}).  Panel (b): $R_{\Delta x_{c}}^{(c)}\left(p_{T}\right)$ for the result of FONLL and these of PYTHIA8 in three multiplicity classes. \label{fig:charm_rdxc}} 
\end{centering}
\end{figure*}

We apply the online calculator of FONLL theoretical method \citep{Cacciari:1998it,Cacciari:2001td}
to calculate $p_{T}$ spectrum of charm quarks in $pp$ collisions
at $\sqrt{s}=$ 13 TeV. The result is shown in Fig.~\ref{fig:charm_rdxc}
(a), where the shadow area denotes the scale uncertainties in calculations.
The solid line is the fit to center values of FONLL calculations by
Eq.~(\ref{eq:fcpt_para}) with parameter values $b=1.3$, $M=6.4$
GeV/c, $n=4.16$, and $c=0.34$ GeV/c. Fig.~\ref{fig:charm_rdxc}(b)
show the calculated $R_{\Delta x_{c}}^{(c)}\left(p_{T}\right)$. We
see that it monotonically increases with $p_{T}$ and the change magnitude
is about two in the range $p_{T}<10$ GeV/c. We also run PYTHIA 8
(version 8.243) \citep{Sjostrand:2014zea} to obtain the $p_{T}$
spectra of charm quarks in three multiplicity classes and calculate
their $R_{\Delta x_{c}}^{(c)}\left(p_{T}\right)$. Here, the low,
intermediate and high multiplicity classes are defined as events within
multiplicity interval {[}1,10{]}, {[}11,19{]} and {[}20,60{]} with
the multiplicity $\left\langle dN_{ch}/d\eta\right\rangle =$ 4.4,
14.3 and 25.0 within the pseudo-rapidity interval $\left|\eta\right|<0.5$,
respectively. Results of $R_{\Delta x_{c}}^{(c)}\left(p_{T}\right)$
are shown in Fig.~\ref{fig:charm_rdxc}(b). We see that $R_{\Delta x_{c}}^{(c)}\left(p_{T}\right)$
calculated by PYTHIA8 has a certain multiplicity dependence and its
$p_{T}$ dependence is similar with that of FONLL calculations. 

\section{The possible property of $R_{corr}^{(cl)}\left(p_{T}\right)$\label{sec:R_corr_cl}}

The term $R_{corr}^{(cl)}\left(p_{T}\right)$ defined in Eq.~(\ref{eq:R_cl_corr})
denotes effects of momentum correlations between different quark flavors
on the $p_{T}$ dependence of $\Lambda_{c}^{+}/D^{0}$ ratio. It contains
two correlation functions $C_{cud}(x_{c}p_{T},x_{u}p_{T},x_{d}p_{T})$
and $C_{c\bar{u}}(x_{c}^{'}p_{T},x_{u}^{'}p_{T})$. In the case of
vanishing momentum correlations $C_{cud}=0$ and $C_{c\bar{l}}=0$,
$R_{corr}^{(cl)}\left(p_{T}\right)=1$ and contribute trivially to
$\Lambda_{c}^{+}/D^{0}$ ratio.

In high energy $pp$ collisions, charm quarks are dominantly produced
by initial hard parton-parton scattering process and possibly interact
with neighboring light-flavor (anti-)quarks before they hadronize.
Part of these interactions is non-perturbative and is difficult to
calculate from first principles. On the other hand, quark momentum
correlations are also difficult to reversely obtained from experimental
data which are mainly of inclusive distributions of particles at present.

In order to roughly understand the possible property of $R_{corr}^{(cl)}\left(p_{T}\right)$
and its influence on $\Lambda_{c}^{+}/D^{0}$ ratio in $pp$ collisions,
we apply the event generator PYTHIA 8 (version 8.243) with default
tunes, as an example, to study the possible property of $R_{corr}^{(cl)}\left(p_{T}\right)$.
We use PYTHIA8 to firstly calculate $f_{c\bar{u}}(p_{T,c},p_{T,\bar{u}},\Delta\phi_{c\bar{u}})$
at $\left|y\right|<0.5$ where $\Delta\phi_{c\bar{u}}$ is the azimuthal
angle difference between $\bar{u}$ and $c$. Then we take the limit
$p_{T,\bar{u}}\to p_{T,c}m_{u}/m_{c}$ and $\Delta\phi_{c\bar{u}}\to0$
for $f_{c\bar{u}}(p_{T,c},p_{T,\bar{u}},\Delta\phi_{c\bar{u}})$ to
obtain the $f_{c\bar{u}}\left(x_{c}^{'}p_{T},x_{u}^{'}p_{T}\right)$
and corresponding $\mathcal{C}_{c\bar{u}}\left(x_{c}^{'}p_{T},x_{u}^{'}p_{T}\right)$
by Eq.~(\ref{eq:fuc_decomp}). 

Because the fraction of the event simultaneously consisting of $u$,
$d$, and $c$ in the mid-rapidity range $\left|y\right|<0.5$ is
very small, it is very hard to directly calculate $f_{cud}\left(x_{c}p_{T},x_{u}p_{T},x_{d}p_{T}\right)$,
in particular, in low multiplicity events. In the general multi-variable
statistics $\left\langle abc\right\rangle =\left\langle a\right\rangle \left\langle b\right\rangle \left\langle c\right\rangle +\left\langle a\right\rangle C_{bc}+\left\langle b\right\rangle C_{ac}+\left\langle c\right\rangle C_{ab}+C_{abc}$,
$C_{ab}$ is standard two-body correlation and $C_{abc}$ is three-body
correlation which is a higher order term compared with two-body correlations.
Here, we neglect the contribution of high-order correlation $C_{cud}$
and obtain 
\begin{align}
    &\mathcal{C}_{udc}\left(x_{u}p_{T},x_{d}p_{T},x_{c}p_{T}\right) \\
    &\approx\mathcal{C}_{ud}\left(x_{u}p_{T},x_{d}p_{T}\right)+\mathcal{C}_{cu}\left(x_{c}p_{T},x_{u}p_{T}\right)+\mathcal{C}_{cd}\left(x_{c}p_{T},x_{d}p_{T}\right),\nonumber
\end{align}
and then we obtain 
\begin{align}
 & R_{corr}^{(cl)}\left(p_{T}\right)\nonumber \\
    & =\frac{1}{1+\mathcal{C}_{c\bar{u}}\left(x_{c}^{'}p_{T},x_{u}^{'}p_{T}\right)} \Big( 1+\mathcal{C}_{ud}\left(x_{u}p_{T},x_{d}p_{T}\right) \nonumber \\
    &\hspace{60pt}+\mathcal{C}_{cu}\left(x_{c}p_{T},x_{u}p_{T}\right) +\mathcal{C}_{cd}\left(x_{c}p_{T},x_{d}p_{T}\right) \Big) \nonumber \\
 & \approx 1+\mathcal{C}_{ud}\left(x_{u}p_{T},x_{d}p_{T}\right)+2\mathcal{C}_{cu}\left(x_{c}p_{T},x_{u}p_{T}\right) \nonumber \\
    &-\mathcal{C}_{c\bar{u}}\left(x_{c}^{'}p_{T},x_{u}^{'}p_{T}\right).
\end{align}
In the second line, we have assumed that $\mathcal{C}_{c\bar{u}}$
is small and $\mathcal{C}_{cu}\approx\mathcal{C}_{cd}$. 

\begin{figure}
\includegraphics[width=0.9\linewidth]{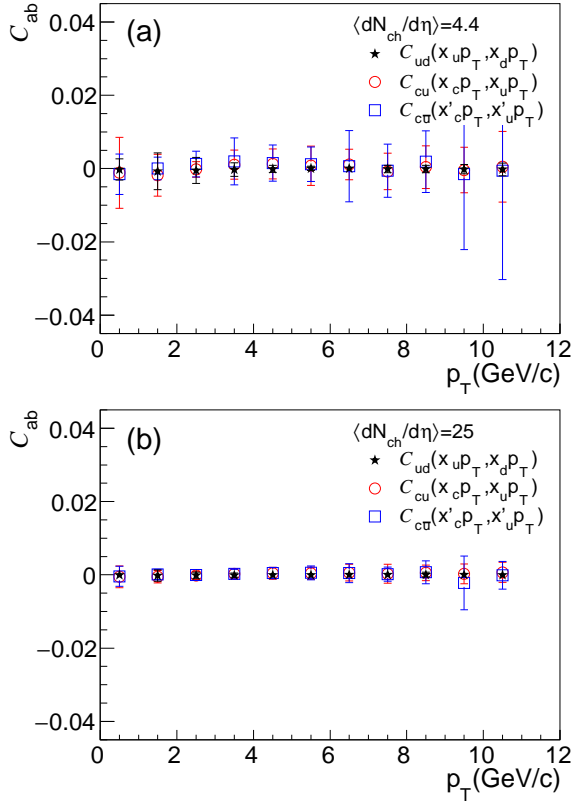}
    \caption{Two quark correlations $\mathcal{C}_{q_{1}q_{2}}$ as the function of $p_{T}$ in low and high multiplicity events calculated by PYTHIA8.  \label{fig:Cab}}
\end{figure}

In Fig.~\ref{fig:Cab}, we show results of two-quark correlation
$\mathcal{C}_{q_{1}q_{2}}$ as the function of $p_{T}$ in low multiplicity
events with $\left\langle dN_{ch}/d\eta\right\rangle =4.4$ and those
in high multiplicity events with $\left\langle dN_{ch}/d\eta\right\rangle =25.0$.
We see that $\mathcal{C}_{ud}$, $\mathcal{C}_{cu}$ and $\mathcal{C}_{c\bar{u}}$
are very small in both low and high multiplicity events. We emphasize
that these statistics only serve as a qualitatively estimation because
of the following reasons. Quarks and antiquarks are only small fractions
of final-state parton system in PYTHIA simulations. Gluons that take
large fraction of the parton system are not involved in current statistics.
Non-perturbative interactions of these quarks and gluons near hadronization
are not fully simulated in PYTHIA8, which may generate some momentum
correlations among quarks and antiquarks of different flavors. In
addition, collectivity or collective flow of partons may be formed
in high multiplicity events \citep{Khachatryan:2016txc,Zhao:2017rgg,Nagle:2018nvi,Bierlich:2017vhg,ALICE:2020ibs},
which may also generate some momentum correlations among quarks and
antiquarks. These possible influence on $R_{corr}^{(cl)}\left(p_{T}\right)$
are left for future study. In this paper, we take $R_{corr}^{(cl)}\left(p_{T}\right)\approx1$
for the moment in the following studies. 

\section{Result of $\Lambda_{c}^{+}/D^{0}$ ratio\label{sec:Ratio_LamC_D0_compare}}

In this section, we combine effects of above four terms to calculate
the $\Lambda_{c}^{+}/D^{0}$ ratio and compare with the latest data
of ALICE collaboration \citep{ALICE:2021npz}. Here, we neglect the
effect of the momentum correlation $R_{corr}^{(cl)}\left(p_{T}\right)$
between charm quarks and light-flavor (anti-)quarks by taking $R_{corr}^{(cl)}\left(p_{T}\right)\approx1$.
This is equivalent to the independence approximation for the joint
momentum distribution $f_{c\bar{l}}\left(p_{T,c},p_{T,\bar{l}}\right)=f_{c}\left(p_{T,c}\right)f_{\bar{l}}\left(p_{T,\bar{l}}\right)$
and $f_{cll'}\left(p_{T,c},p_{T,l},p_{T,l'}\right)=f_{c}\left(p_{T,c}\right)f_{l}\left(p_{T,l}\right)f_{l'}\left(p_{T,l'}\right)$.
In this case, 
\begin{equation}
\frac{f_{\Lambda_{c}}(p_{T})}{f_{D_{0}}(p_{T})}\approx R^{(N_{q_{i}})}\left[\frac{A_{\Lambda_{c}}}{A_{D_{0}}}R_{\Delta x_{c}}^{(c)}\left(p_{T}\right)R_{\Delta x_{l}}^{(l)}\left(p_{T}\right)\right]
\end{equation}
where the property of $R^{(N_{q_{i}})}$, $R_{\Delta x_{c}}^{(c)}\left(p_{T}\right)$
and $R_{\Delta x_{l}}^{(l)}\left(p_{T}\right)$ are already shown
in Eqs. (\ref{eq:R_Nqi_final}), (\ref{eq:R_dxl_approx}) and (\ref{eq:R_dxlc_appro}).
Coefficients $A_{D_{0}}$ and $A_{\Lambda_{c}}$ can be calculated
by Eqs.~(\ref{eq:A_Mcl_def}) and (\ref{eq:A_Bcll_def}) with the
inclusive distributions of charm and light-flavor quarks shown in
Figs.~\ref{fig:fu_Rl_dxl} and \ref{fig:charm_rdxc}. 

In order to study the multiplicity dependence of the $\Lambda_{c}^{+}/D^{0}$
ratio, we take the light-flavor quark spectrum $f_{l}(p_{T})$ in
event class $\left\langle dN_{ch}/d\eta\right\rangle =25.75$ (class
I) and $\left\langle dN_{ch}/d\eta\right\rangle =4.64$ (class IX)
as examples of high multiplicity events and low multiplicity events,
respectively. For charm quark $p_{T}$ spectrum $f_{c}(p_{T})$, we
have two choices. One is that of FONLL calculation and we use it in
both high multiplicity events and low multiplicity events. Another
is that of PYTHIA8 calculation where in Sec.\ref{sec:R_dt_c} we have
obtained $f_{c}(p_{T})$ in high multiplicity events $\left\langle dN_{ch}/d\eta\right\rangle =25.0$
and that in low multiplicity events $\left\langle dN_{ch}/d\eta\right\rangle =4.4$.
The term $R^{(N_{q_{i}})}$ contains two multiplicity-dependent factors
$\lambda_{s}$ and $R_{B/M}^{(c)}$. As shown by Eq.~(\ref{eq:R_Nqi_final}),
strangeness factor $\lambda_{s}$ causes little changes on $\Lambda_{c}^{+}/D^{0}$
ratio and we can safely neglect its multiplicity dependence. Baryon-to-meson
production competition factor $R_{B/M}^{(c)}$ is a relatively-free
parameter which cannot be fixed in our model at present. In Sec. \ref{sec:Rnq},
we have discussed its possible multiplicity dependence.

\begin{figure*}[!htpb]
\includegraphics[scale=0.4]{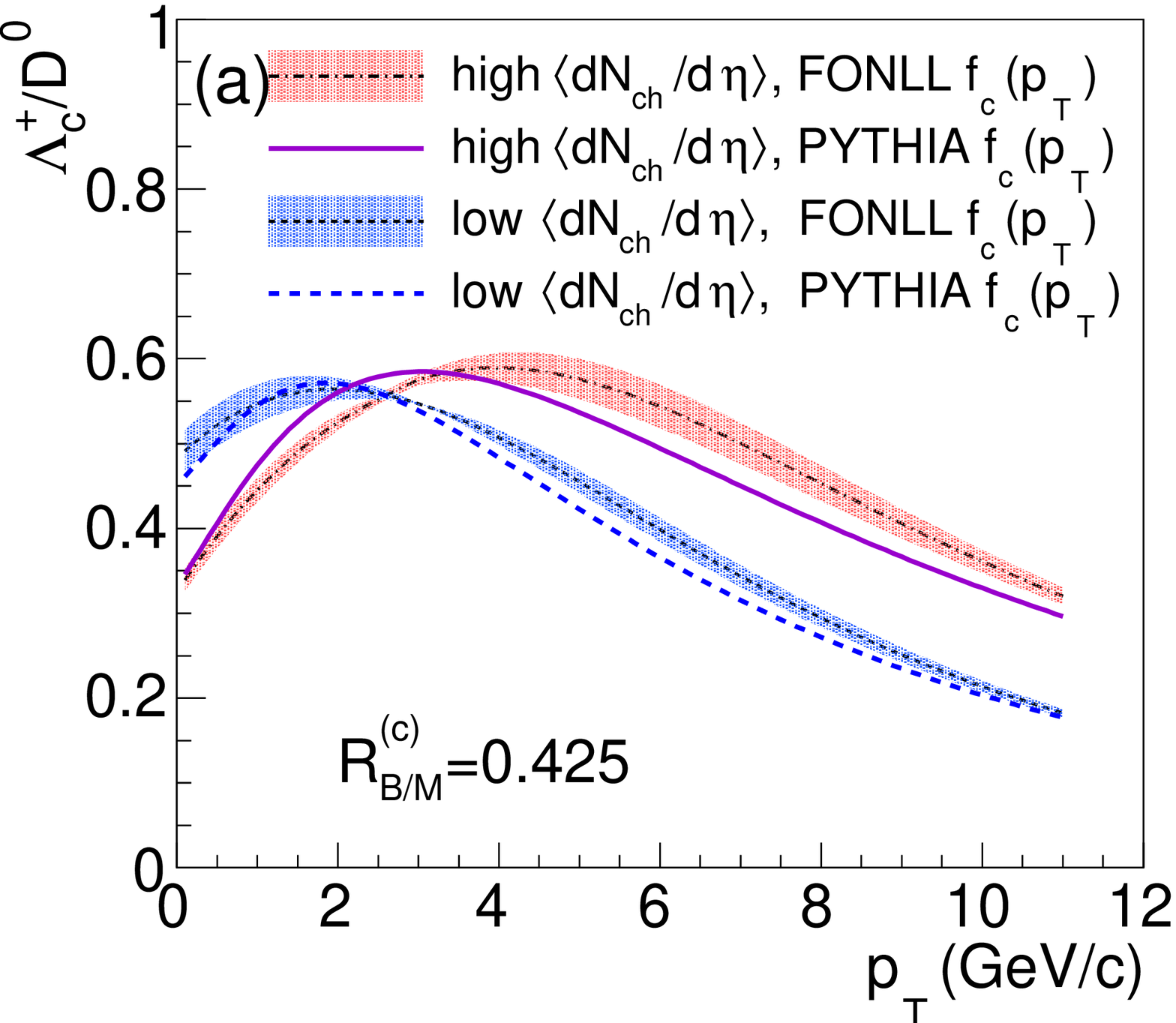}\includegraphics[scale=0.4]{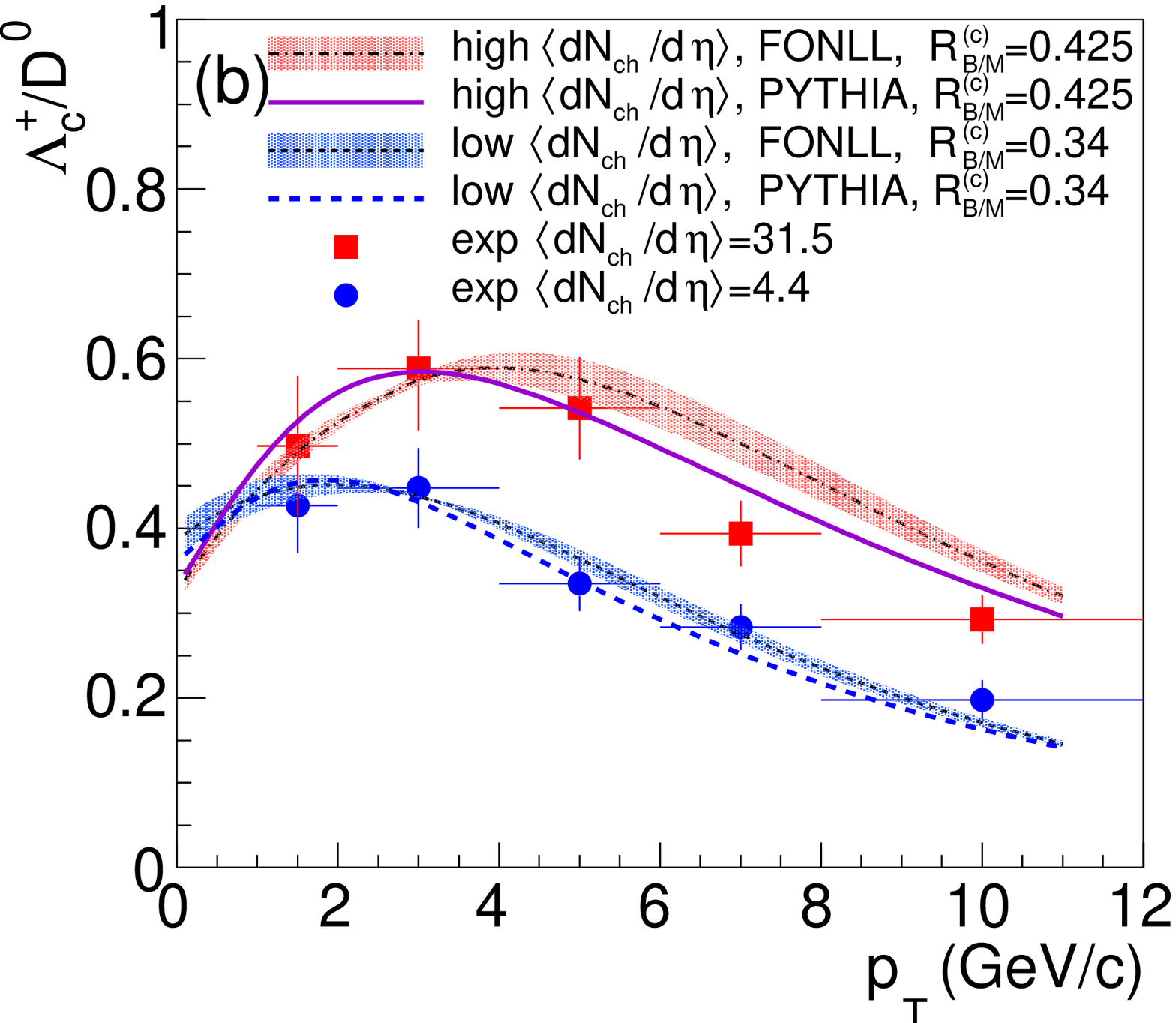}
    \caption{$\Lambda_{c}^{+}/D^{0}$ ratio as the function of $p_{T}$ in low and high multiplicity events in $pp$ collisions at $\sqrt{s}=13$ TeV (a) and the comparison with experimental data \citep{ALICE:2021npz} (b). See text for the detailed discussions. \label{fig:Ratio_LamC_D0_final}}
\end{figure*}

In Fig. \ref{fig:Ratio_LamC_D0_final}(a), we show results of $\Lambda_{c}^{+}/D^{0}$
ratio as the function of $p_{T}$ in high multiplicity events and
low multiplicity events at given $R_{B/M}^{(c)}=0.425$. The
purpose of taking the same $R_{B/M}^{(c)}$ is to let us focus on
the influence of $p_{T}$ spectra of light-flavor quarks and charm
quarks on the $p_{T}$ dependence of $\Lambda_{c}^{+}/D^{0}$ ratio.
Here, high multiplicity corresponds to $\left\langle dN_{ch}/d\eta\right\rangle \approx25$
since we combine effect of $f_{l}(p_{T})$ in event class $\left\langle dN_{ch}/d\eta\right\rangle =25.75$
(class I) and that of $f_{c}(p_{T})$ in events with $\left\langle dN_{ch}/d\eta\right\rangle =25$
in PYTHIA8 calculations. Low multiplicity corresponds to $\left\langle dN_{ch}/d\eta\right\rangle \approx4.5$
since we combine effect of $f_{l}(p_{T})$ in event class $\left\langle dN_{ch}/d\eta\right\rangle =4.64$
(class IX) and that of $f_{c}(p_{T})$ in events with $\left\langle dN_{ch}/d\eta\right\rangle =4.4$
in PYTHIA8 calculations. The dot-dashed line with shadow band is the
result when $f_{c}(p_{T})$ calculated by FONLL and $f_{l}(p_{T})$
in high multiplicity events are used. The dotted line with shadow
band is the result when $f_{c}(p_{T})$ calculated by FONLL and $f_{l}(p_{T})$
in low multiplicity events are used. Comparing two results, we see
an obvious multiplicity dependence of $\Lambda_{c}^{+}/D^{0}$ ratio
caused by that of light-flavor quark spectrum. As $p_{T}\gtrsim3$
GeV/c, the $\Lambda_{c}^{+}/D^{0}$ ratio in high multiplicity events
is obviously higher than that in low multiplicity events. This is
because $f_{l}(p_{T})$ in high multiplicity events is obviously flatter
than that in low multiplicity events, see Fig.~\ref{fig:fu_Rl_dxl}(a).
As $p_{T}\lesssim3$ GeV/c, the situation is reversed. This is because
at the same $R_{B/M}^{(c)}$ the ratio of $p_{T}$-integrated yield
of $\Lambda_{c}^{+}$ to $D^{0}$ is almost the same, the suppress
of $\Lambda_{c}^{+}$ at $p_{T}\gtrsim3$ GeV/c in low multiplicity
events is offset by the enhancement of $\Lambda_{c}^{+}$ at low $p_{T}$.
We also show the result of $\Lambda_{c}^{+}/D^{0}$ ratio when the
$f_{c}(p_{T})$ calculated by PYTHIA8 in high multiplicity events
is applied. The result, the solid line, is higher than that with FONLL
calculated $f_{c}(p_{T})$ in low $p_{T}$ range and is slightly smaller
than the latter as $p_{T}\gtrsim3$ GeV/c. In the range $3\lesssim p_{T}\lesssim8$
GeV/c, the result of $\Lambda_{c}^{+}/D^{0}$ ratio with $f_{c}(p_{T})$
calculated by PYTHIA8 in low multiplicity events, the dashed line,
is also slightly smaller than that with FONLL calculated $f_{c}(p_{T})$.
Comparing results using FONLL calculated $f_{c}(p_{T})$ with those
using PYTHIA8 calculated $f_{c}(p_{T})$, we see a weak influence
of $f_{c}(p_{T})$ uncertainty on $\Lambda_{c}^{+}/D^{0}$ ratio. 

In addition, we see that $\Lambda_{c}^{+}/D^{0}$ ratios in our model
in both high and low multiplicity events always exhibit a non-monotonic
$p_{T}$ dependence. This is mainly because of the property of $R_{\Delta x_{l}}^{(l)}\left(p_{T}\right)$
in Eq.~(\ref{eq:R_dxl_approx}), which is determined by the property
of $p_{T}$ spectrum of light-flavor quark as shown in Fig.~\ref{fig:fu_Rl_dxl}(b)
and (c). The term $R_{\Delta x_{c}}^{(c)}\left(p_{T}\right)$, as
shown in Fig.~\ref{fig:charm_rdxc}(b), strengthen the increase of
the $\Lambda_{c}^{+}/D^{0}$ ratio in the low $p_{T}$ range ($p_{T}\lesssim2-3$
GeV/c) but offsets the decrease of the $R_{\Delta x_{l}}^{(l)}\left(p_{T}\right)$
as $p_{T}\gtrsim3$ GeV/c to a certain extent and leads to the relatively
weak decrease of the $\Lambda_{c}^{+}/D^{0}$ ratio as $p_{T}\gtrsim3$
GeV/c. 

In Fig.~\ref{fig:Ratio_LamC_D0_final}(b), we test above theoretical
results by experimental data of $\Lambda_{c}^{+}/D^{0}$ ratio in
$pp$ collisions at $\sqrt{s}=13$ TeV \citep{ALICE:2021npz}. Here,
we show data of $\Lambda_{c}^{+}/D^{0}$ ratio in V0M multiplicity
class with $\left\langle dN_{ch}/d\eta=31.5\right\rangle $ as an
example of high multiplicity events and data in V0M multiplicity class
with $\left\langle dN_{ch}/d\eta=4.4\right\rangle $ as an example
of low multiplicity events. Because the light-flavor quark $p_{T}$
spectrum used in our calculation is extracted from experimental data
of light-flavor hadrons in V0M event class I with $\left\langle dN_{ch}/d\eta\right\rangle =25.75$
and class IX with $\left\langle dN_{ch}/d\eta\right\rangle =4.64$,
the underlying events of our theoretical calculations are not exactly
same as those for data of $\Lambda_{c}^{+}/D^{0}$ ratio but the difference
should be small due to the similar multiplicity. In high multiplicity
events, we see that theoretical result with FONLL calculated $f_{c}(p_{T})$,
the dot-dashed line, is consistent with the first two datum points
with $p_{T}<3$ GeV/c and is higher than data at larger $p_{T}$ to
a certain extent. Theoretical result with PYTHIA8 calculated $f_{c}(p_{T})$
in high multiplicity events, the solid line, is more close to experimental
data in comparison with that with FONLL calculated $f_{c}(p_{T})$.
Here, the baryon-to-meson competition factor $R_{B/M}^{(c)}$ is taken
to be 0.425, the possible saturation value in high multiplicity events. 

According to discussions in Sec.~\ref{sec:Rnq}, we expect a suppression
of $R_{B/M}^{(c)}$ in low multiplicity events and a rough estimation
based on Poisson distribution of quark numbers gives about 20\% suppression.
Therefore, here we use $R_{B/M}^{(c)}=0.34$ to calculate the $\Lambda_{c}^{+}/D^{0}$
ratio in low multiplicity events. We see a good agreement with experimental
data in the available $p_{T}$ range.

In above comparison with experimental data, we have considered the
decay contributions of other single-charm hadrons to $\Lambda_{c}^{+}/D^{0}$
ratio by the yield term $R^{(N_{q_{i}})}$ Eq.~(\ref{eq:R_Nqi_final}).
The decay influence on the shape of $\Lambda_{c}^{+}/D^{0}$ ratio (as the function of $p_{T}$)
is numerically studied and is found to be quite small and therefore the comparison with experimental data
in Fig.~\ref{fig:Ratio_LamC_D0_final}(b) is little changed. 

\begin{figure*}
\includegraphics[width=0.75\linewidth]{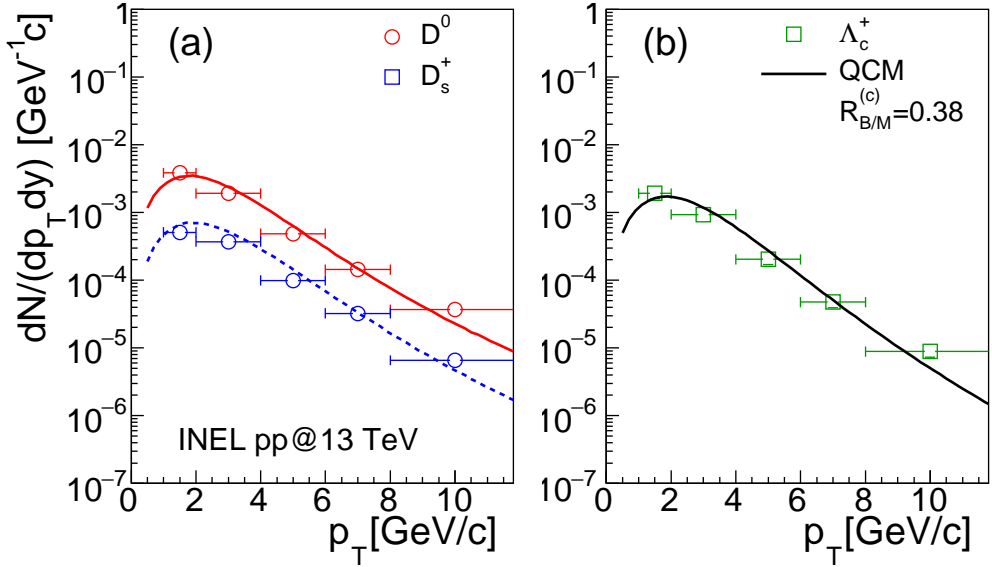}
    \caption{$p_{T}$ spectra of $D^{0}$, $D_{s}^{+}$ (a) and $\Lambda_{c}^{+}$ (b) in inelastic $pp$ collisions at $\sqrt{s}=13$ TeV. Lines are model results and symbols are experimental data \citep{ALICE:2021npz}.\label{fig:fptcharm_inel}}
\end{figure*}

In Fig. \ref{fig:fptcharm_inel}, we show $p_{T}$ spectra of $D^{0}$,
$D_{s}^{+}$ and $\Lambda_{c}^{+}$ in inelastic $pp$ collisions
at $\sqrt{s}=13$ TeV and compare them with experimental data \citep{ALICE:2021npz}.
Here, $p_{T}$ spectra of light-flavor quarks in inelastic events
have been obtained in Ref.~\citep{Zhang:2018vyr} and $p_{T}$ spectrum
of charm quarks is taken from FONLL calculations shown in Fig.~\ref{fig:charm_rdxc}
(a) with a $p_{T}$-integrated yield density $dN_{c}/dy=0.025$. The
model parameter $R_{B/M}^{(c)}$ is set to be 0.38, which falls in
between the value in high multiplicity events and that in low multiplicity
events. The decay contributions from other single-charm hadrons in
ground state are systematically considered. We see that experimental
data of $D^{0}$, $D_{s}^{+}$ and $\Lambda_{c}^{+}$ are self-consistently
explained in our model. 

\section{Summary\label{sec:summary}}

We have applied an equal-velocity quark combination model to study
the $\Lambda_{c}^{+}/D^{0}$ ratio as the function of $p_{T}$ in
$pp$ collisions at $\sqrt{s}=13$ TeV. Taking advantage of the analytic
feature of the model, we decomposed the $\Lambda_{c}^{+}/D^{0}$ ratio
into four parts and studied their individual influence on the ratio.
Finally, we combined effects of these parts to calculate $\Lambda_{c}^{+}/D^{0}$
ratio as the function of $p_{T}$ and compared theoretical results
with experimental data of $\Lambda_{c}^{+}/D^{0}$ ratio in high and
low multiplicity events in $pp$ collisions at $\sqrt{s}=13$ TeV. 

The first part of $\Lambda_{c}^{+}/D^{0}$ decomposition is the term
of light-flavor quark numbers. We summarized this part into a compact
form containing a strangeness factor  and a baryon-to-meson competition
factor defined in our model. The multiplicity dependence of strangeness
has very weak influence on the $\Lambda_{c}^{+}/D^{0}$ ratio. The
global effect of light-flavor quark numbers is manifested by the baryon-to-meson
competition factor. In low multiplicity events where quark numbers
are relatively small, charm quark has relatively small chance to interact
with two light-flavor quarks and thus the production of charm baryon
should be suppressed to a certain extent. We adopted a Poisson distribution
as an example to roughly estimate this suppression and found about
20\% suppression of the $\Lambda_{c}^{+}/D^{0}$ ratio in low multiplicity
events with mid-rapidity $\left\langle dN_{ch}/d\eta\right\rangle \approx4.5$
in comparison with that in high multiplicity events with $\left\langle dN_{ch}/d\eta\right\rangle \approx25$. 

The second part of $\Lambda_{c}^{+}/D^{0}$ decomposition is the term
containing the normalized $p_{T}$ spectrum of up/down quarks. Considering
the small difference between momentum fraction ($x_{u}$) of up/down
quark in $\Lambda_{c}^{+}$ and that ($x_{u}^{'}$) in $D^{0}$, we
adopted the Taylor expansion method and reduced this part to the inclusive
distribution of up/down quarks $f_{u}^{(n)}\left((2x_{u}-x_{u}^{'})p_{T}\right)$
with a good numerical accuracy. This suggests that the shape of up
quark $p_{T}$ spectrum directly transmit to the $p_{T}$ dependence
of the $\Lambda_{c}^{+}/D^{0}$ ratio. In particular, $p_{T}$ spectrum
of up/down quarks in high multiplicity events behaves a thermal like
distribution in the low $p_{T}$ range, which will lead to an obviously
non-monotonic $p_{T}$ dependence of $\Lambda_{c}^{+}/D^{0}$ ratio
in the low and intermediate $p_{T}$ range. 

The third part of $\Lambda_{c}^{+}/D^{0}$ decomposition is the term
containing the normalized $p_{T}$ spectrum of charm quarks. We adopted
FONLL method and PYTHIA8 to calculate charm quark $p_{T}$ spectrum,
respectively. This part increases with $p_{T}$ and therefore weaken
the effect of the second part. We found that results of PYTHIA8 are
roughly close to that of FONLL calculations but have a weak multiplicity
dependence. The fourth part of $\Lambda_{c}^{+}/D^{0}$ decomposition
is the term containing momentum correlation between up/down quarks
and charm quarks. PYTHIA8 simulation on two-quark correlations show
a negligible momentum correlation between up/down quarks and charm
quarks. Therefore, we neglect the effect of this part in calculation
of $\Lambda_{c}^{+}/D^{0}$ ratio in this paper. 

Finally, we combined these parts and calculate $\Lambda_{c}^{+}/D^{0}$
ratio as the function of $p_{T}$ in $pp$ collisions at $\sqrt{s}=13$
TeV. In calculations, we chosen the $p_{T}$ spectrum of up/down quarks
in multiplicity class IX with $\left\langle dN_{ch}/d\eta\right\rangle =4.64$
as an example of low multiplicity events and that in multiplicity
class I with $\left\langle dN_{ch}/d\eta\right\rangle =25.75$ as
an example of high multiplicity events. $p_{T}$ spectrum of charm
quarks from PYTHIA8 calculation with $\left\langle dN_{ch}/d\eta\right\rangle =4.4$
and that with $\left\langle dN_{ch}/d\eta\right\rangle =25$ are taken
in order to be consistent with the selected $p_{T}$ spectrum of light-flavor
quarks at similar $\left\langle dN_{ch}/d\eta\right\rangle $. Finally,
we obtained $\Lambda_{c}^{+}/D^{0}$ ratio as the function of $p_{T}$
in high multiplicity events with $\left\langle dN_{ch}/d\eta\right\rangle \approx25$
and that in low multiplicity events with $\left\langle dN_{ch}/d\eta\right\rangle \approx4.5$.
We compared our theoretical results with experimental data of $\Lambda_{c}^{+}/D^{0}$
ratio in multiplicity class with $\left\langle dN_{ch}/d\eta\right\rangle =31.5$
and those with $\left\langle dN_{ch}/d\eta\right\rangle =4.4$. In
high multiplicity events with a baryon-to-meson competition factor
$R_{B/M}^{(c)}=0.425$, we found that theoretical result with PYTHIA8
calculated charm $p_{T}$ spectrum is in better agreement with experiment
data than that with FONLL calculated charm $p_{T}$ spectrum. In low
multiplicity events with a suppressed $R_{B/M}^{(c)}=0.34$, theoretical
result with FONLL calculated charm $p_{T}$ spectrum and that with
PYTHIA8 calculated charm $p_{T}$ spectrum are both in good agreement
with experimental data. 

$R_{B/M}^{(c)}$ is a dynamical parameter of our model which denotes the competition between the formation of charm baryon and that of charm meson at a charm quark hadronization and can not be predicted by the model of present version. 
Based on the present work and our previous studies \cite{Song:2018tpv,Li:2017zuj,Li:2021nhq,Song:2021ojn}, we found that $R_{B/M}^{(c)}$ is about $0.385-0.425$ in inelastic and relatively-high multiplicity events $pp$ and $p$Pb collisions and is suppressed in small multiplicity events but it may also increase to a certain extent in AA collisions \cite{Wang:2019fcg}. This possibly multiplicity-dependent property of $R_{B/M}^{(c)}$ is an interesting property of charm quark hadronization, which is not seen in the hadronization of light-flavor quarks where we found an relatively stable baryon-to-meson production competition \cite{Shao:2017eok}. Its explanation is beyond the  model of current version and should introduce further dynamic considerations.  For example, by considering the possible production of some $cll'$ resonances which is predicted by quark model but is not observed yet in experiments \cite{He:2019vgs,Chen:2020drg}, the hadronization channel of charm quark to charm baryons should increase which may prompt the possible increase of $R_{B/M}^{(c)}$ in high multiplicity events where those resonances will have more practical chance to produce. Also, by considering the competition of quark combination mechanism and quark fragmentation mechanism and, in particular, the multiplicity dependence of such competition \cite{Plumari:2017ntm,Minissale:2020bif}, the effective $R_{B/M}^{(c)}$ can also be changed in our model. These possible dynamics at charm quark hadronization are deserved to study in the future to improve our model. 

\section{Acknowledgments}

This work is supported by Shandong Provincial Natural Science
Foundation (Grants No. ZR2019YQ06 and No. ZR2019MA053), and Higher
Educational Youth Innovation Science and Technology Program of Shandong
Province (Grant No. 2019KJJ010).

\bibliographystyle{apsrev4-1}
\bibliography{ref}

\begin{thebibliography}{45}%
\makeatletter
\providecommand \@ifxundefined [1]{%
 \@ifx{#1\undefined}
}%
\providecommand \@ifnum [1]{%
 \ifnum #1\expandafter \@firstoftwo
 \else \expandafter \@secondoftwo
 \fi
}%
\providecommand \@ifx [1]{%
 \ifx #1\expandafter \@firstoftwo
 \else \expandafter \@secondoftwo
 \fi
}%
\providecommand \natexlab [1]{#1}%
\providecommand \enquote  [1]{``#1''}%
\providecommand \bibnamefont  [1]{#1}%
\providecommand \bibfnamefont [1]{#1}%
\providecommand \citenamefont [1]{#1}%
\providecommand \href@noop [0]{\@secondoftwo}%
\providecommand \href [0]{\begingroup \@sanitize@url \@href}%
\providecommand \@href[1]{\@@startlink{#1}\@@href}%
\providecommand \@@href[1]{\endgroup#1\@@endlink}%
\providecommand \@sanitize@url [0]{\catcode `\\12\catcode `\$12\catcode
  `\&12\catcode `\#12\catcode `\^12\catcode `\_12\catcode `\%12\relax}%
\providecommand \@@startlink[1]{}%
\providecommand \@@endlink[0]{}%
\providecommand \url  [0]{\begingroup\@sanitize@url \@url }%
\providecommand \@url [1]{\endgroup\@href {#1}{\urlprefix }}%
\providecommand \urlprefix  [0]{URL }%
\providecommand \Eprint [0]{\href }%
\providecommand \doibase [0]{http://dx.doi.org/}%
\providecommand \selectlanguage [0]{\@gobble}%
\providecommand \bibinfo  [0]{\@secondoftwo}%
\providecommand \bibfield  [0]{\@secondoftwo}%
\providecommand \translation [1]{[#1]}%
\providecommand \BibitemOpen [0]{}%
\providecommand \bibitemStop [0]{}%
\providecommand \bibitemNoStop [0]{.\EOS\space}%
\providecommand \EOS [0]{\spacefactor3000\relax}%
\providecommand \BibitemShut  [1]{\csname bibitem#1\endcsname}%
\let\auto@bib@innerbib\@empty
\bibitem [{\citenamefont {Acharya}\ \emph
  {et~al.}(2021{\natexlab{a}})\citenamefont {Acharya} \emph
  {et~al.}}]{ALICE:2020jsh}%
  \BibitemOpen
  \bibfield  {author} {\bibinfo {author} {\bibfnamefont {S.}~\bibnamefont
  {Acharya}} \emph {et~al.} (\bibinfo {collaboration} {ALICE}),\ }\href
  {\doibase 10.1140/epjc/s10052-020-08690-5} {\bibfield  {journal} {\bibinfo
  {journal} {Eur. Phys. J. C}\ }\textbf {\bibinfo {volume} {81}},\ \bibinfo
  {pages} {256} (\bibinfo {year} {2021}{\natexlab{a}})},\ \Eprint
  {http://arxiv.org/abs/2005.11120} {arXiv:2005.11120 [nucl-ex]} \BibitemShut
  {NoStop}%
\bibitem [{\citenamefont {Khachatryan}\ \emph {et~al.}(2011)\citenamefont
  {Khachatryan} \emph {et~al.}}]{Khachatryan:2011tm}%
  \BibitemOpen
  \bibfield  {author} {\bibinfo {author} {\bibfnamefont {V.}~\bibnamefont
  {Khachatryan}} \emph {et~al.} (\bibinfo {collaboration} {CMS}),\ }\href
  {\doibase 10.1007/JHEP05(2011)064} {\bibfield  {journal} {\bibinfo  {journal}
  {JHEP}\ }\textbf {\bibinfo {volume} {05}},\ \bibinfo {pages} {064} (\bibinfo
  {year} {2011})},\ \Eprint {http://arxiv.org/abs/1102.4282} {arXiv:1102.4282
  [hep-ex]} \BibitemShut {NoStop}%
\bibitem [{\citenamefont {Adam}\ \emph {et~al.}(2017)\citenamefont {Adam} \emph
  {et~al.}}]{ALICE:2017jyt}%
  \BibitemOpen
  \bibfield  {author} {\bibinfo {author} {\bibfnamefont {J.}~\bibnamefont
  {Adam}} \emph {et~al.} (\bibinfo {collaboration} {ALICE}),\ }\href {\doibase
  10.1038/nphys4111} {\bibfield  {journal} {\bibinfo  {journal} {Nature Phys.}\
  }\textbf {\bibinfo {volume} {13}},\ \bibinfo {pages} {535} (\bibinfo {year}
  {2017})},\ \Eprint {http://arxiv.org/abs/1606.07424} {arXiv:1606.07424
  [nucl-ex]} \BibitemShut {NoStop}%
\bibitem [{\citenamefont {Song}\ \emph {et~al.}(2017)\citenamefont {Song},
  \citenamefont {Gou}, \citenamefont {Shao},\ and\ \citenamefont
  {Liang}}]{Song:2017gcz}%
  \BibitemOpen
  \bibfield  {author} {\bibinfo {author} {\bibfnamefont {J.}~\bibnamefont
  {Song}}, \bibinfo {author} {\bibfnamefont {X.-r.}\ \bibnamefont {Gou}},
  \bibinfo {author} {\bibfnamefont {F.-l.}\ \bibnamefont {Shao}}, \ and\
  \bibinfo {author} {\bibfnamefont {Z.-T.}\ \bibnamefont {Liang}},\ }\href
  {\doibase 10.1016/j.physletb.2017.10.012} {\bibfield  {journal} {\bibinfo
  {journal} {Phys. Lett.}\ }\textbf {\bibinfo {volume} {B774}},\ \bibinfo
  {pages} {516} (\bibinfo {year} {2017})},\ \Eprint
  {http://arxiv.org/abs/1707.03949} {arXiv:1707.03949 [hep-ph]} \BibitemShut
  {NoStop}%
\bibitem [{\citenamefont {Zhang}\ \emph {et~al.}(2020)\citenamefont {Zhang},
  \citenamefont {Li}, \citenamefont {Shao},\ and\ \citenamefont
  {Song}}]{Zhang:2018vyr}%
  \BibitemOpen
  \bibfield  {author} {\bibinfo {author} {\bibfnamefont {J.-w.}\ \bibnamefont
  {Zhang}}, \bibinfo {author} {\bibfnamefont {H.-h.}\ \bibnamefont {Li}},
  \bibinfo {author} {\bibfnamefont {F.-l.}\ \bibnamefont {Shao}}, \ and\
  \bibinfo {author} {\bibfnamefont {J.}~\bibnamefont {Song}},\ }\href {\doibase
  10.1088/1674-1137/44/1/014101} {\bibfield  {journal} {\bibinfo  {journal}
  {Chin. Phys.}\ }\textbf {\bibinfo {volume} {C44}},\ \bibinfo {pages} {014101}
  (\bibinfo {year} {2020})},\ \Eprint {http://arxiv.org/abs/1811.00975}
  {arXiv:1811.00975 [hep-ph]} \BibitemShut {NoStop}%
\bibitem [{\citenamefont {Li}\ \emph {et~al.}(2021)\citenamefont {Li},
  \citenamefont {Shao},\ and\ \citenamefont {Song}}]{Li:2021nhq}%
  \BibitemOpen
  \bibfield  {author} {\bibinfo {author} {\bibfnamefont {H.-h.}\ \bibnamefont
  {Li}}, \bibinfo {author} {\bibfnamefont {F.-l.}\ \bibnamefont {Shao}}, \ and\
  \bibinfo {author} {\bibfnamefont {J.}~\bibnamefont {Song}},\ }\href {\doibase
  10.1088/1674-1137/ac1ef9} {\bibfield  {journal} {\bibinfo  {journal} {Chin.
  Phys. C}\ }\textbf {\bibinfo {volume} {45}},\ \bibinfo {pages} {113105}
  (\bibinfo {year} {2021})},\ \Eprint {http://arxiv.org/abs/2103.14900}
  {arXiv:2103.14900 [hep-ph]} \BibitemShut {NoStop}%
\bibitem [{\citenamefont {Acharya}\ \emph {et~al.}(2018)\citenamefont {Acharya}
  \emph {et~al.}}]{Acharya:2017kfy}%
  \BibitemOpen
  \bibfield  {author} {\bibinfo {author} {\bibfnamefont {S.}~\bibnamefont
  {Acharya}} \emph {et~al.} (\bibinfo {collaboration} {ALICE}),\ }\href
  {\doibase 10.1007/JHEP04(2018)108} {\bibfield  {journal} {\bibinfo  {journal}
  {JHEP}\ }\textbf {\bibinfo {volume} {04}},\ \bibinfo {pages} {108} (\bibinfo
  {year} {2018})},\ \Eprint {http://arxiv.org/abs/1712.09581} {arXiv:1712.09581
  [nucl-ex]} \BibitemShut {NoStop}%
\bibitem [{\citenamefont {Sirunyan}\ \emph {et~al.}(2020)\citenamefont
  {Sirunyan} \emph {et~al.}}]{CMS:2019uws}%
  \BibitemOpen
  \bibfield  {author} {\bibinfo {author} {\bibfnamefont {A.~M.}\ \bibnamefont
  {Sirunyan}} \emph {et~al.} (\bibinfo {collaboration} {CMS}),\ }\href
  {\doibase 10.1016/j.physletb.2020.135328} {\bibfield  {journal} {\bibinfo
  {journal} {Phys. Lett. B}\ }\textbf {\bibinfo {volume} {803}},\ \bibinfo
  {pages} {135328} (\bibinfo {year} {2020})},\ \Eprint
  {http://arxiv.org/abs/1906.03322} {arXiv:1906.03322 [hep-ex]} \BibitemShut
  {NoStop}%
\bibitem [{\citenamefont {Acharya}\ \emph
  {et~al.}(2021{\natexlab{b}})\citenamefont {Acharya} \emph
  {et~al.}}]{ALICE:2020wla}%
  \BibitemOpen
  \bibfield  {author} {\bibinfo {author} {\bibfnamefont {S.}~\bibnamefont
  {Acharya}} \emph {et~al.} (\bibinfo {collaboration} {ALICE}),\ }\href
  {\doibase 10.1103/PhysRevC.104.054905} {\bibfield  {journal} {\bibinfo
  {journal} {Phys. Rev. C}\ }\textbf {\bibinfo {volume} {104}},\ \bibinfo
  {pages} {054905} (\bibinfo {year} {2021}{\natexlab{b}})},\ \Eprint
  {http://arxiv.org/abs/2011.06079} {arXiv:2011.06079 [nucl-ex]} \BibitemShut
  {NoStop}%
\bibitem [{\citenamefont {Acharya}\ \emph
  {et~al.}(2022{\natexlab{a}})\citenamefont {Acharya} \emph
  {et~al.}}]{ALICE:2021rzj}%
  \BibitemOpen
  \bibfield  {author} {\bibinfo {author} {\bibfnamefont {S.}~\bibnamefont
  {Acharya}} \emph {et~al.} (\bibinfo {collaboration} {ALICE}),\ }\href
  {\doibase 10.1103/PhysRevLett.128.012001} {\bibfield  {journal} {\bibinfo
  {journal} {Phys. Rev. Lett.}\ }\textbf {\bibinfo {volume} {128}},\ \bibinfo
  {pages} {012001} (\bibinfo {year} {2022}{\natexlab{a}})},\ \Eprint
  {http://arxiv.org/abs/2106.08278} {arXiv:2106.08278 [hep-ex]} \BibitemShut
  {NoStop}%
\bibitem [{\citenamefont {Ortiz~Velasquez}\ \emph {et~al.}(2013)\citenamefont
  {Ortiz~Velasquez}, \citenamefont {Christiansen}, \citenamefont
  {Cuautle~Flores}, \citenamefont {Maldonado~Cervantes},\ and\ \citenamefont
  {Pai{\'c}}}]{Ortiz:2013yxa}%
  \BibitemOpen
  \bibfield  {author} {\bibinfo {author} {\bibfnamefont {A.}~\bibnamefont
  {Ortiz~Velasquez}}, \bibinfo {author} {\bibfnamefont {P.}~\bibnamefont
  {Christiansen}}, \bibinfo {author} {\bibfnamefont {E.}~\bibnamefont
  {Cuautle~Flores}}, \bibinfo {author} {\bibfnamefont {I.}~\bibnamefont
  {Maldonado~Cervantes}}, \ and\ \bibinfo {author} {\bibfnamefont
  {G.}~\bibnamefont {Pai{\'c}}},\ }\href {\doibase
  10.1103/PhysRevLett.111.042001} {\bibfield  {journal} {\bibinfo  {journal}
  {Phys. Rev. Lett.}\ }\textbf {\bibinfo {volume} {111}},\ \bibinfo {pages}
  {042001} (\bibinfo {year} {2013})},\ \Eprint {http://arxiv.org/abs/1303.6326}
  {arXiv:1303.6326 [hep-ph]} \BibitemShut {NoStop}%
\bibitem [{\citenamefont {Bierlich}\ \emph {et~al.}(2015)\citenamefont
  {Bierlich}, \citenamefont {Gustafson}, \citenamefont {L{\"o}nnblad},\ and\
  \citenamefont {Tarasov}}]{Bierlich:2014xba}%
  \BibitemOpen
  \bibfield  {author} {\bibinfo {author} {\bibfnamefont {C.}~\bibnamefont
  {Bierlich}}, \bibinfo {author} {\bibfnamefont {G.}~\bibnamefont {Gustafson}},
  \bibinfo {author} {\bibfnamefont {L.}~\bibnamefont {L{\"o}nnblad}}, \ and\
  \bibinfo {author} {\bibfnamefont {A.}~\bibnamefont {Tarasov}},\ }\href
  {\doibase 10.1007/JHEP03(2015)148} {\bibfield  {journal} {\bibinfo  {journal}
  {JHEP}\ }\textbf {\bibinfo {volume} {03}},\ \bibinfo {pages} {148} (\bibinfo
  {year} {2015})},\ \Eprint {http://arxiv.org/abs/1412.6259} {arXiv:1412.6259
  [hep-ph]} \BibitemShut {NoStop}%
\bibitem [{\citenamefont {Bierlich}\ and\ \citenamefont
  {Christiansen}(2015)}]{Bierlich:2015rha}%
  \BibitemOpen
  \bibfield  {author} {\bibinfo {author} {\bibfnamefont {C.}~\bibnamefont
  {Bierlich}}\ and\ \bibinfo {author} {\bibfnamefont {J.~R.}\ \bibnamefont
  {Christiansen}},\ }\href {\doibase 10.1103/PhysRevD.92.094010} {\bibfield
  {journal} {\bibinfo  {journal} {Phys. Rev.}\ }\textbf {\bibinfo {volume}
  {D92}},\ \bibinfo {pages} {094010} (\bibinfo {year} {2015})},\ \Eprint
  {http://arxiv.org/abs/1507.02091} {arXiv:1507.02091 [hep-ph]} \BibitemShut
  {NoStop}%
\bibitem [{\citenamefont {Gieseke}\ \emph {et~al.}(2018)\citenamefont
  {Gieseke}, \citenamefont {Kirchgae\ss{}er},\ and\ \citenamefont
  {Pl\"atzer}}]{Gieseke:2017clv}%
  \BibitemOpen
  \bibfield  {author} {\bibinfo {author} {\bibfnamefont {S.}~\bibnamefont
  {Gieseke}}, \bibinfo {author} {\bibfnamefont {P.}~\bibnamefont
  {Kirchgae\ss{}er}}, \ and\ \bibinfo {author} {\bibfnamefont {S.}~\bibnamefont
  {Pl\"atzer}},\ }\href {\doibase 10.1140/epjc/s10052-018-5585-7} {\bibfield
  {journal} {\bibinfo  {journal} {Eur. Phys. J. C}\ }\textbf {\bibinfo {volume}
  {78}},\ \bibinfo {pages} {99} (\bibinfo {year} {2018})},\ \Eprint
  {http://arxiv.org/abs/1710.10906} {arXiv:1710.10906 [hep-ph]} \BibitemShut
  {NoStop}%
\bibitem [{\citenamefont {Christiansen}\ and\ \citenamefont
  {Skands}(2015)}]{Christiansen:2015yqa}%
  \BibitemOpen
  \bibfield  {author} {\bibinfo {author} {\bibfnamefont {J.~R.}\ \bibnamefont
  {Christiansen}}\ and\ \bibinfo {author} {\bibfnamefont {P.~Z.}\ \bibnamefont
  {Skands}},\ }\href {\doibase 10.1007/JHEP08(2015)003} {\bibfield  {journal}
  {\bibinfo  {journal} {JHEP}\ }\textbf {\bibinfo {volume} {08}},\ \bibinfo
  {pages} {003} (\bibinfo {year} {2015})},\ \Eprint
  {http://arxiv.org/abs/1505.01681} {arXiv:1505.01681 [hep-ph]} \BibitemShut
  {NoStop}%
\bibitem [{\citenamefont {Fischer}\ and\ \citenamefont
  {Sj{\"o}strand}(2017)}]{Fischer:2016zzs}%
  \BibitemOpen
  \bibfield  {author} {\bibinfo {author} {\bibfnamefont {N.}~\bibnamefont
  {Fischer}}\ and\ \bibinfo {author} {\bibfnamefont {T.}~\bibnamefont
  {Sj{\"o}strand}},\ }\href {\doibase 10.1007/JHEP01(2017)140} {\bibfield
  {journal} {\bibinfo  {journal} {JHEP}\ }\textbf {\bibinfo {volume} {01}},\
  \bibinfo {pages} {140} (\bibinfo {year} {2017})},\ \Eprint
  {http://arxiv.org/abs/1610.09818} {arXiv:1610.09818 [hep-ph]} \BibitemShut
  {NoStop}%
\bibitem [{\citenamefont {Sj\"ostrand}(2018)}]{Sjostrand:2017cdm}%
  \BibitemOpen
  \bibfield  {author} {\bibinfo {author} {\bibfnamefont {T.}~\bibnamefont
  {Sj\"ostrand}},\ }\href {\doibase 10.1142/9789813227767_0010} {\bibfield
  {journal} {\bibinfo  {journal} {Adv. Ser. Direct. High Energy Phys.}\
  }\textbf {\bibinfo {volume} {29}},\ \bibinfo {pages} {191} (\bibinfo {year}
  {2018})},\ \Eprint {http://arxiv.org/abs/1706.02166} {arXiv:1706.02166
  [hep-ph]} \BibitemShut {NoStop}%
\bibitem [{\citenamefont {Acharya}\ \emph
  {et~al.}(2022{\natexlab{b}})\citenamefont {Acharya} \emph
  {et~al.}}]{ALICE:2021npz}%
  \BibitemOpen
  \bibfield  {author} {\bibinfo {author} {\bibfnamefont {S.}~\bibnamefont
  {Acharya}} \emph {et~al.} (\bibinfo {collaboration} {ALICE}),\ }\href
  {\doibase 10.1016/j.physletb.2022.137065} {\bibfield  {journal} {\bibinfo
  {journal} {Phys. Lett. B}\ }\textbf {\bibinfo {volume} {829}},\ \bibinfo
  {pages} {137065} (\bibinfo {year} {2022}{\natexlab{b}})},\ \Eprint
  {http://arxiv.org/abs/2111.11948} {arXiv:2111.11948 [nucl-ex]} \BibitemShut
  {NoStop}%
\bibitem [{\citenamefont {Wilkinson}(2022)}]{Wilkinson:2022tyl}%
  \BibitemOpen
  \bibfield  {author} {\bibinfo {author} {\bibfnamefont {J.}~\bibnamefont
  {Wilkinson}} (\bibinfo {collaboration} {ALICE}),\ }in\ \href@noop {} {\emph
  {\bibinfo {booktitle} {{56th Rencontres de Moriond~ on QCD and High Energy
  Interactions~}}}}\ (\bibinfo {year} {2022})\ \Eprint
  {http://arxiv.org/abs/2205.07196} {arXiv:2205.07196 [hep-ex]} \BibitemShut
  {NoStop}%
\bibitem [{\citenamefont {ALICE}(2022)}]{ALICE:2022ych}%
  \BibitemOpen
  \bibfield  {author} {\bibinfo {author} {\bibnamefont {ALICE}},\ }\href@noop
  {} {\enquote {\bibinfo {title} {{First measurement of
  $\Lambda_\mathrm{c}^{+}$ production down to $p_\mathrm{T} = 0$ in pp and p-Pb
  collisions at $\sqrt{s_\mathrm{NN}} = 5.02$ TeV}},}\ } (\bibinfo {year}
  {2022}),\ \Eprint {http://arxiv.org/abs/2211.14032} {arXiv:2211.14032
  [nucl-ex]} \BibitemShut {NoStop}%
\bibitem [{\citenamefont {He}\ and\ \citenamefont {Rapp}(2019)}]{He:2019tik}%
  \BibitemOpen
  \bibfield  {author} {\bibinfo {author} {\bibfnamefont {M.}~\bibnamefont
  {He}}\ and\ \bibinfo {author} {\bibfnamefont {R.}~\bibnamefont {Rapp}},\
  }\href {\doibase 10.1016/j.physletb.2019.06.004} {\bibfield  {journal}
  {\bibinfo  {journal} {Phys. Lett. B}\ }\textbf {\bibinfo {volume} {795}},\
  \bibinfo {pages} {117} (\bibinfo {year} {2019})},\ \Eprint
  {http://arxiv.org/abs/1902.08889} {arXiv:1902.08889 [nucl-th]} \BibitemShut
  {NoStop}%
\bibitem [{\citenamefont {Minissale}\ \emph {et~al.}(2021)\citenamefont
  {Minissale}, \citenamefont {Plumari},\ and\ \citenamefont
  {Greco}}]{Minissale:2020bif}%
  \BibitemOpen
  \bibfield  {author} {\bibinfo {author} {\bibfnamefont {V.}~\bibnamefont
  {Minissale}}, \bibinfo {author} {\bibfnamefont {S.}~\bibnamefont {Plumari}},
  \ and\ \bibinfo {author} {\bibfnamefont {V.}~\bibnamefont {Greco}},\ }\href
  {\doibase 10.1016/j.physletb.2021.136622} {\bibfield  {journal} {\bibinfo
  {journal} {Phys. Lett. B}\ }\textbf {\bibinfo {volume} {821}},\ \bibinfo
  {pages} {136622} (\bibinfo {year} {2021})},\ \Eprint
  {http://arxiv.org/abs/2012.12001} {arXiv:2012.12001 [hep-ph]} \BibitemShut
  {NoStop}%
\bibitem [{\citenamefont {Chen}\ and\ \citenamefont {He}(2021)}]{Chen:2020drg}%
  \BibitemOpen
  \bibfield  {author} {\bibinfo {author} {\bibfnamefont {Y.}~\bibnamefont
  {Chen}}\ and\ \bibinfo {author} {\bibfnamefont {M.}~\bibnamefont {He}},\
  }\href {\doibase 10.1016/j.physletb.2021.136144} {\bibfield  {journal}
  {\bibinfo  {journal} {Phys. Lett. B}\ }\textbf {\bibinfo {volume} {815}},\
  \bibinfo {pages} {136144} (\bibinfo {year} {2021})},\ \Eprint
  {http://arxiv.org/abs/2011.14328} {arXiv:2011.14328 [hep-ph]} \BibitemShut
  {NoStop}%
\bibitem [{\citenamefont {Song}\ \emph {et~al.}(2018)\citenamefont {Song},
  \citenamefont {Li},\ and\ \citenamefont {Shao}}]{Song:2018tpv}%
  \BibitemOpen
  \bibfield  {author} {\bibinfo {author} {\bibfnamefont {J.}~\bibnamefont
  {Song}}, \bibinfo {author} {\bibfnamefont {H.-h.}\ \bibnamefont {Li}}, \ and\
  \bibinfo {author} {\bibfnamefont {F.-l.}\ \bibnamefont {Shao}},\ }\href
  {\doibase 10.1140/epjc/s10052-018-5817-x} {\bibfield  {journal} {\bibinfo
  {journal} {Eur. Phys. J.}\ }\textbf {\bibinfo {volume} {C78}},\ \bibinfo
  {pages} {344} (\bibinfo {year} {2018})},\ \Eprint
  {http://arxiv.org/abs/1801.09402} {arXiv:1801.09402 [hep-ph]} \BibitemShut
  {NoStop}%
\bibitem [{\citenamefont {Li}\ \emph {et~al.}(2018)\citenamefont {Li},
  \citenamefont {Shao}, \citenamefont {Song},\ and\ \citenamefont
  {Wang}}]{Li:2017zuj}%
  \BibitemOpen
  \bibfield  {author} {\bibinfo {author} {\bibfnamefont {H.-H.}\ \bibnamefont
  {Li}}, \bibinfo {author} {\bibfnamefont {F.-L.}\ \bibnamefont {Shao}},
  \bibinfo {author} {\bibfnamefont {J.}~\bibnamefont {Song}}, \ and\ \bibinfo
  {author} {\bibfnamefont {R.-Q.}\ \bibnamefont {Wang}},\ }\href {\doibase
  10.1103/PhysRevC.97.064915} {\bibfield  {journal} {\bibinfo  {journal} {Phys.
  Rev.}\ }\textbf {\bibinfo {volume} {C97}},\ \bibinfo {pages} {064915}
  (\bibinfo {year} {2018})},\ \Eprint {http://arxiv.org/abs/1712.08921}
  {arXiv:1712.08921 [hep-ph]} \BibitemShut {NoStop}%
\bibitem [{\citenamefont {Gou}\ \emph {et~al.}(2017)\citenamefont {Gou},
  \citenamefont {Shao}, \citenamefont {Wang}, \citenamefont {Li},\ and\
  \citenamefont {Song}}]{Gou:2017foe}%
  \BibitemOpen
  \bibfield  {author} {\bibinfo {author} {\bibfnamefont {X.-r.}\ \bibnamefont
  {Gou}}, \bibinfo {author} {\bibfnamefont {F.-l.}\ \bibnamefont {Shao}},
  \bibinfo {author} {\bibfnamefont {R.-q.}\ \bibnamefont {Wang}}, \bibinfo
  {author} {\bibfnamefont {H.-h.}\ \bibnamefont {Li}}, \ and\ \bibinfo {author}
  {\bibfnamefont {J.}~\bibnamefont {Song}},\ }\href {\doibase
  10.1103/PhysRevD.96.094010} {\bibfield  {journal} {\bibinfo  {journal} {Phys.
  Rev.}\ }\textbf {\bibinfo {volume} {D96}},\ \bibinfo {pages} {094010}
  (\bibinfo {year} {2017})},\ \Eprint {http://arxiv.org/abs/1707.06906}
  {arXiv:1707.06906 [hep-ph]} \BibitemShut {NoStop}%
\bibitem [{\citenamefont {Song}\ \emph
  {et~al.}(2021{\natexlab{a}})\citenamefont {Song}, \citenamefont {Li},\ and\
  \citenamefont {Shao}}]{Song:2021ygg}%
  \BibitemOpen
  \bibfield  {author} {\bibinfo {author} {\bibfnamefont {J.}~\bibnamefont
  {Song}}, \bibinfo {author} {\bibfnamefont {H.-h.}\ \bibnamefont {Li}}, \ and\
  \bibinfo {author} {\bibfnamefont {F.-l.}\ \bibnamefont {Shao}},\ }\href
  {\doibase 10.1140/epjc/s10052-020-08760-8} {\bibfield  {journal} {\bibinfo
  {journal} {Eur. Phys. J. C}\ }\textbf {\bibinfo {volume} {81}},\ \bibinfo
  {pages} {1} (\bibinfo {year} {2021}{\natexlab{a}})},\ \Eprint
  {http://arxiv.org/abs/2008.03017} {arXiv:2008.03017 [nucl-th]} \BibitemShut
  {NoStop}%
\bibitem [{\citenamefont {Song}\ \emph {et~al.}(2022)\citenamefont {Song},
  \citenamefont {Li},\ and\ \citenamefont {Shao}}]{Song:2021ojn}%
  \BibitemOpen
  \bibfield  {author} {\bibinfo {author} {\bibfnamefont {J.}~\bibnamefont
  {Song}}, \bibinfo {author} {\bibfnamefont {H.-h.}\ \bibnamefont {Li}}, \ and\
  \bibinfo {author} {\bibfnamefont {F.-l.}\ \bibnamefont {Shao}},\ }\href
  {\doibase 10.1103/PhysRevD.105.074027} {\bibfield  {journal} {\bibinfo
  {journal} {Phys. Rev. D}\ }\textbf {\bibinfo {volume} {105}},\ \bibinfo
  {pages} {074027} (\bibinfo {year} {2022})},\ \Eprint
  {http://arxiv.org/abs/2109.11722} {arXiv:2109.11722 [hep-ph]} \BibitemShut
  {NoStop}%
\bibitem [{\citenamefont {Song}\ \emph
  {et~al.}(2021{\natexlab{b}})\citenamefont {Song}, \citenamefont {Wang},
  \citenamefont {Li}, \citenamefont {Wang},\ and\ \citenamefont
  {Shao}}]{Song:2020kak}%
  \BibitemOpen
  \bibfield  {author} {\bibinfo {author} {\bibfnamefont {J.}~\bibnamefont
  {Song}}, \bibinfo {author} {\bibfnamefont {X.-F.}\ \bibnamefont {Wang}},
  \bibinfo {author} {\bibfnamefont {H.-H.}\ \bibnamefont {Li}}, \bibinfo
  {author} {\bibfnamefont {R.-Q.}\ \bibnamefont {Wang}}, \ and\ \bibinfo
  {author} {\bibfnamefont {F.-L.}\ \bibnamefont {Shao}},\ }\href {\doibase
  10.1103/PhysRevC.103.034907} {\bibfield  {journal} {\bibinfo  {journal}
  {Phys. Rev. C}\ }\textbf {\bibinfo {volume} {103}},\ \bibinfo {pages}
  {034907} (\bibinfo {year} {2021}{\natexlab{b}})},\ \Eprint
  {http://arxiv.org/abs/2007.14588} {arXiv:2007.14588 [nucl-th]} \BibitemShut
  {NoStop}%
\bibitem [{\citenamefont {Wang}\ \emph {et~al.}(2020)\citenamefont {Wang},
  \citenamefont {Song}, \citenamefont {Shao},\ and\ \citenamefont
  {Liang}}]{Wang:2019fcg}%
  \BibitemOpen
  \bibfield  {author} {\bibinfo {author} {\bibfnamefont {R.-Q.}\ \bibnamefont
  {Wang}}, \bibinfo {author} {\bibfnamefont {J.}~\bibnamefont {Song}}, \bibinfo
  {author} {\bibfnamefont {F.-L.}\ \bibnamefont {Shao}}, \ and\ \bibinfo
  {author} {\bibfnamefont {Z.-T.}\ \bibnamefont {Liang}},\ }\href {\doibase
  10.1103/PhysRevC.101.054903} {\bibfield  {journal} {\bibinfo  {journal}
  {Phys. Rev. C}\ }\textbf {\bibinfo {volume} {101}},\ \bibinfo {pages}
  {054903} (\bibinfo {year} {2020})},\ \Eprint
  {http://arxiv.org/abs/1911.00823} {arXiv:1911.00823 [hep-ph]} \BibitemShut
  {NoStop}%
\bibitem [{\citenamefont {Ding}\ \emph {et~al.}(2015)\citenamefont {Ding},
  \citenamefont {Mukherjee}, \citenamefont {Ohno}, \citenamefont {Petreczky},\
  and\ \citenamefont {Schadler}}]{Ding:2015fca}%
  \BibitemOpen
  \bibfield  {author} {\bibinfo {author} {\bibfnamefont {H.~T.}\ \bibnamefont
  {Ding}}, \bibinfo {author} {\bibfnamefont {S.}~\bibnamefont {Mukherjee}},
  \bibinfo {author} {\bibfnamefont {H.}~\bibnamefont {Ohno}}, \bibinfo {author}
  {\bibfnamefont {P.}~\bibnamefont {Petreczky}}, \ and\ \bibinfo {author}
  {\bibfnamefont {H.~P.}\ \bibnamefont {Schadler}},\ }\href {\doibase
  10.1103/PhysRevD.92.074043} {\bibfield  {journal} {\bibinfo  {journal} {Phys.
  Rev. D}\ }\textbf {\bibinfo {volume} {92}},\ \bibinfo {pages} {074043}
  (\bibinfo {year} {2015})},\ \Eprint {http://arxiv.org/abs/1507.06637}
  {arXiv:1507.06637 [hep-lat]} \BibitemShut {NoStop}%
\bibitem [{\citenamefont {Olive}\ \emph {et~al.}(2014)\citenamefont {Olive}
  \emph {et~al.}}]{Agashe:2014kda}%
  \BibitemOpen
  \bibfield  {author} {\bibinfo {author} {\bibfnamefont {K.~A.}\ \bibnamefont
  {Olive}} \emph {et~al.} (\bibinfo {collaboration} {Particle Data Group}),\
  }\href {\doibase 10.1088/1674-1137/38/9/090001} {\bibfield  {journal}
  {\bibinfo  {journal} {Chin. Phys. C}\ }\textbf {\bibinfo {volume} {38}},\
  \bibinfo {pages} {090001} (\bibinfo {year} {2014})}\BibitemShut {NoStop}%
\bibitem [{\citenamefont {Acharya}\ \emph {et~al.}(2019)\citenamefont {Acharya}
  \emph {et~al.}}]{Acharya:2019mgn}%
  \BibitemOpen
  \bibfield  {author} {\bibinfo {author} {\bibfnamefont {S.}~\bibnamefont
  {Acharya}} \emph {et~al.} (\bibinfo {collaboration} {ALICE}),\ }\href
  {\doibase 10.1140/epjc/s10052-019-6873-6} {\bibfield  {journal} {\bibinfo
  {journal} {Eur. Phys. J. C}\ }\textbf {\bibinfo {volume} {79}},\ \bibinfo
  {pages} {388} (\bibinfo {year} {2019})},\ \Eprint
  {http://arxiv.org/abs/1901.07979} {arXiv:1901.07979 [nucl-ex]} \BibitemShut
  {NoStop}%
\bibitem [{\citenamefont {Acharya}\ \emph
  {et~al.}(2020{\natexlab{a}})\citenamefont {Acharya} \emph
  {et~al.}}]{ALICE:2019avo}%
  \BibitemOpen
  \bibfield  {author} {\bibinfo {author} {\bibfnamefont {S.}~\bibnamefont
  {Acharya}} \emph {et~al.} (\bibinfo {collaboration} {ALICE}),\ }\href
  {\doibase 10.1140/epjc/s10052-020-7673-8} {\bibfield  {journal} {\bibinfo
  {journal} {Eur. Phys. J. C}\ }\textbf {\bibinfo {volume} {80}},\ \bibinfo
  {pages} {167} (\bibinfo {year} {2020}{\natexlab{a}})},\ \Eprint
  {http://arxiv.org/abs/1908.01861} {arXiv:1908.01861 [nucl-ex]} \BibitemShut
  {NoStop}%
\bibitem [{\citenamefont {Cacciari}\ \emph {et~al.}(1998)\citenamefont
  {Cacciari}, \citenamefont {Greco},\ and\ \citenamefont
  {Nason}}]{Cacciari:1998it}%
  \BibitemOpen
  \bibfield  {author} {\bibinfo {author} {\bibfnamefont {M.}~\bibnamefont
  {Cacciari}}, \bibinfo {author} {\bibfnamefont {M.}~\bibnamefont {Greco}}, \
  and\ \bibinfo {author} {\bibfnamefont {P.}~\bibnamefont {Nason}},\ }\href
  {\doibase 10.1088/1126-6708/1998/05/007} {\bibfield  {journal} {\bibinfo
  {journal} {JHEP}\ }\textbf {\bibinfo {volume} {05}},\ \bibinfo {pages} {007}
  (\bibinfo {year} {1998})},\ \Eprint {http://arxiv.org/abs/hep-ph/9803400}
  {arXiv:hep-ph/9803400} \BibitemShut {NoStop}%
\bibitem [{\citenamefont {Cacciari}\ \emph {et~al.}(2001)\citenamefont
  {Cacciari}, \citenamefont {Frixione},\ and\ \citenamefont
  {Nason}}]{Cacciari:2001td}%
  \BibitemOpen
  \bibfield  {author} {\bibinfo {author} {\bibfnamefont {M.}~\bibnamefont
  {Cacciari}}, \bibinfo {author} {\bibfnamefont {S.}~\bibnamefont {Frixione}},
  \ and\ \bibinfo {author} {\bibfnamefont {P.}~\bibnamefont {Nason}},\ }\href
  {\doibase 10.1088/1126-6708/2001/03/006} {\bibfield  {journal} {\bibinfo
  {journal} {JHEP}\ }\textbf {\bibinfo {volume} {03}},\ \bibinfo {pages} {006}
  (\bibinfo {year} {2001})},\ \Eprint {http://arxiv.org/abs/hep-ph/0102134}
  {arXiv:hep-ph/0102134} \BibitemShut {NoStop}%
\bibitem [{\citenamefont {Sj\"ostrand}\ \emph {et~al.}(2015)\citenamefont
  {Sj\"ostrand}, \citenamefont {Ask}, \citenamefont {Christiansen},
  \citenamefont {Corke}, \citenamefont {Desai}, \citenamefont {Ilten},
  \citenamefont {Mrenna}, \citenamefont {Prestel}, \citenamefont {Rasmussen},\
  and\ \citenamefont {Skands}}]{Sjostrand:2014zea}%
  \BibitemOpen
  \bibfield  {author} {\bibinfo {author} {\bibfnamefont {T.}~\bibnamefont
  {Sj\"ostrand}}, \bibinfo {author} {\bibfnamefont {S.}~\bibnamefont {Ask}},
  \bibinfo {author} {\bibfnamefont {J.~R.}\ \bibnamefont {Christiansen}},
  \bibinfo {author} {\bibfnamefont {R.}~\bibnamefont {Corke}}, \bibinfo
  {author} {\bibfnamefont {N.}~\bibnamefont {Desai}}, \bibinfo {author}
  {\bibfnamefont {P.}~\bibnamefont {Ilten}}, \bibinfo {author} {\bibfnamefont
  {S.}~\bibnamefont {Mrenna}}, \bibinfo {author} {\bibfnamefont
  {S.}~\bibnamefont {Prestel}}, \bibinfo {author} {\bibfnamefont {C.~O.}\
  \bibnamefont {Rasmussen}}, \ and\ \bibinfo {author} {\bibfnamefont {P.~Z.}\
  \bibnamefont {Skands}},\ }\href {\doibase 10.1016/j.cpc.2015.01.024}
  {\bibfield  {journal} {\bibinfo  {journal} {Comput. Phys. Commun.}\ }\textbf
  {\bibinfo {volume} {191}},\ \bibinfo {pages} {159} (\bibinfo {year}
  {2015})},\ \Eprint {http://arxiv.org/abs/1410.3012} {arXiv:1410.3012
  [hep-ph]} \BibitemShut {NoStop}%
\bibitem [{\citenamefont {Khachatryan}\ \emph {et~al.}(2017)\citenamefont
  {Khachatryan} \emph {et~al.}}]{Khachatryan:2016txc}%
  \BibitemOpen
  \bibfield  {author} {\bibinfo {author} {\bibfnamefont {V.}~\bibnamefont
  {Khachatryan}} \emph {et~al.} (\bibinfo {collaboration} {CMS}),\ }\href
  {\doibase 10.1016/j.physletb.2016.12.009} {\bibfield  {journal} {\bibinfo
  {journal} {Phys. Lett.}\ }\textbf {\bibinfo {volume} {B765}},\ \bibinfo
  {pages} {193} (\bibinfo {year} {2017})},\ \Eprint
  {http://arxiv.org/abs/1606.06198} {arXiv:1606.06198 [nucl-ex]} \BibitemShut
  {NoStop}%
\bibitem [{\citenamefont {Zhao}\ \emph {et~al.}(2018)\citenamefont {Zhao},
  \citenamefont {Zhou}, \citenamefont {Xu}, \citenamefont {Deng},\ and\
  \citenamefont {Song}}]{Zhao:2017rgg}%
  \BibitemOpen
  \bibfield  {author} {\bibinfo {author} {\bibfnamefont {W.}~\bibnamefont
  {Zhao}}, \bibinfo {author} {\bibfnamefont {Y.}~\bibnamefont {Zhou}}, \bibinfo
  {author} {\bibfnamefont {H.}~\bibnamefont {Xu}}, \bibinfo {author}
  {\bibfnamefont {W.}~\bibnamefont {Deng}}, \ and\ \bibinfo {author}
  {\bibfnamefont {H.}~\bibnamefont {Song}},\ }\href {\doibase
  10.1016/j.physletb.2018.03.022} {\bibfield  {journal} {\bibinfo  {journal}
  {Phys. Lett. B}\ }\textbf {\bibinfo {volume} {780}},\ \bibinfo {pages} {495}
  (\bibinfo {year} {2018})},\ \Eprint {http://arxiv.org/abs/1801.00271}
  {arXiv:1801.00271 [nucl-th]} \BibitemShut {NoStop}%
\bibitem [{\citenamefont {Nagle}\ and\ \citenamefont
  {Zajc}(2018)}]{Nagle:2018nvi}%
  \BibitemOpen
  \bibfield  {author} {\bibinfo {author} {\bibfnamefont {J.~L.}\ \bibnamefont
  {Nagle}}\ and\ \bibinfo {author} {\bibfnamefont {W.~A.}\ \bibnamefont
  {Zajc}},\ }\href {\doibase 10.1146/annurev-nucl-101916-123209} {\bibfield
  {journal} {\bibinfo  {journal} {Ann. Rev. Nucl. Part. Sci.}\ }\textbf
  {\bibinfo {volume} {68}},\ \bibinfo {pages} {211} (\bibinfo {year} {2018})},\
  \Eprint {http://arxiv.org/abs/1801.03477} {arXiv:1801.03477 [nucl-ex]}
  \BibitemShut {NoStop}%
\bibitem [{\citenamefont {Bierlich}\ \emph {et~al.}(2018)\citenamefont
  {Bierlich}, \citenamefont {Gustafson},\ and\ \citenamefont
  {L\"onnblad}}]{Bierlich:2017vhg}%
  \BibitemOpen
  \bibfield  {author} {\bibinfo {author} {\bibfnamefont {C.}~\bibnamefont
  {Bierlich}}, \bibinfo {author} {\bibfnamefont {G.}~\bibnamefont {Gustafson}},
  \ and\ \bibinfo {author} {\bibfnamefont {L.}~\bibnamefont {L\"onnblad}},\
  }\href {\doibase 10.1016/j.physletb.2018.01.069} {\bibfield  {journal}
  {\bibinfo  {journal} {Phys. Lett. B}\ }\textbf {\bibinfo {volume} {779}},\
  \bibinfo {pages} {58} (\bibinfo {year} {2018})},\ \Eprint
  {http://arxiv.org/abs/1710.09725} {arXiv:1710.09725 [hep-ph]} \BibitemShut
  {NoStop}%
\bibitem [{\citenamefont {Acharya}\ \emph
  {et~al.}(2020{\natexlab{b}})\citenamefont {Acharya} \emph
  {et~al.}}]{ALICE:2020ibs}%
  \BibitemOpen
  \bibfield  {author} {\bibinfo {author} {\bibfnamefont {S.}~\bibnamefont
  {Acharya}} \emph {et~al.} (\bibinfo {collaboration} {ALICE}),\ }\href
  {\doibase 10.1016/j.physletb.2020.135849} {\bibfield  {journal} {\bibinfo
  {journal} {Phys. Lett. B}\ }\textbf {\bibinfo {volume} {811}},\ \bibinfo
  {pages} {135849} (\bibinfo {year} {2020}{\natexlab{b}})},\ \Eprint
  {http://arxiv.org/abs/2004.08018} {arXiv:2004.08018 [nucl-ex]} \BibitemShut
  {NoStop}%
\bibitem [{\citenamefont {Shao}\ \emph {et~al.}(2017)\citenamefont {Shao},
  \citenamefont {Wang}, \citenamefont {Wang}, \citenamefont {Li},\ and\
  \citenamefont {Song}}]{Shao:2017eok}%
  \BibitemOpen
  \bibfield  {author} {\bibinfo {author} {\bibfnamefont {F.-l.}\ \bibnamefont
  {Shao}}, \bibinfo {author} {\bibfnamefont {G.-j.}\ \bibnamefont {Wang}},
  \bibinfo {author} {\bibfnamefont {R.-q.}\ \bibnamefont {Wang}}, \bibinfo
  {author} {\bibfnamefont {H.-h.}\ \bibnamefont {Li}}, \ and\ \bibinfo {author}
  {\bibfnamefont {J.}~\bibnamefont {Song}},\ }\href {\doibase
  10.1103/PhysRevC.95.064911} {\bibfield  {journal} {\bibinfo  {journal}
  {Phys.\ Rev.\ C}\ }\textbf {\bibinfo {volume} {95}},\ \bibinfo {pages}
  {064911} (\bibinfo {year} {2017})},\ \Eprint
  {http://arxiv.org/abs/1703.05862} {arXiv:1703.05862 [hep-ph]} \BibitemShut
  {NoStop}%
\bibitem [{\citenamefont {He}\ and\ \citenamefont {Rapp}(2020)}]{He:2019vgs}%
  \BibitemOpen
  \bibfield  {author} {\bibinfo {author} {\bibfnamefont {M.}~\bibnamefont
  {He}}\ and\ \bibinfo {author} {\bibfnamefont {R.}~\bibnamefont {Rapp}},\
  }\href {\doibase 10.1103/PhysRevLett.124.042301} {\bibfield  {journal}
  {\bibinfo  {journal} {Phys. Rev. Lett.}\ }\textbf {\bibinfo {volume} {124}},\
  \bibinfo {pages} {042301} (\bibinfo {year} {2020})},\ \Eprint
  {http://arxiv.org/abs/1905.09216} {arXiv:1905.09216 [nucl-th]} \BibitemShut
  {NoStop}%
\bibitem [{\citenamefont {Plumari}\ \emph {et~al.}(2018)\citenamefont
  {Plumari}, \citenamefont {Minissale}, \citenamefont {Das}, \citenamefont
  {Coci},\ and\ \citenamefont {Greco}}]{Plumari:2017ntm}%
  \BibitemOpen
  \bibfield  {author} {\bibinfo {author} {\bibfnamefont {S.}~\bibnamefont
  {Plumari}}, \bibinfo {author} {\bibfnamefont {V.}~\bibnamefont {Minissale}},
  \bibinfo {author} {\bibfnamefont {S.~K.}\ \bibnamefont {Das}}, \bibinfo
  {author} {\bibfnamefont {G.}~\bibnamefont {Coci}}, \ and\ \bibinfo {author}
  {\bibfnamefont {V.}~\bibnamefont {Greco}},\ }\href {\doibase
  10.1140/epjc/s10052-018-5828-7} {\bibfield  {journal} {\bibinfo  {journal}
  {Eur. Phys. J. C}\ }\textbf {\bibinfo {volume} {78}},\ \bibinfo {pages} {348}
  (\bibinfo {year} {2018})},\ \Eprint {http://arxiv.org/abs/1712.00730}
  {arXiv:1712.00730 [hep-ph]} \BibitemShut {NoStop}%
\end{thebibliography}%

\end{document}